\documentclass[aps,prx,reprint,twocolumn,showpacs,floatfix,longbibliography]{revtex4-2}
\usepackage{mathrsfs,braket}
\usepackage{amssymb, amsbsy, amsmath, latexsym, dsfont, array, layout, graphicx, mathrsfs, color, bm}
\usepackage[normalem]{ulem}
\usepackage{amsthm}
\usepackage[colorlinks=true,linkcolor=red,anchorcolor=red,citecolor=blue, urlcolor=blue]{hyperref}
\usepackage{multirow} 
\usepackage{tikz}
\usepackage{xcolor}
\usepackage{mathrsfs}

\newcommand{\labeledsquare}[1]{%
    \begin{tikzpicture}[baseline={(current bounding box.center)}, scale=0.6]
        \fill[blue!30] (0,0) rectangle (0.8,0.8); 
        \draw (0,0) rectangle (0.8,0.8); 
    \end{tikzpicture}%
}

\newcommand{\labeledrhombus}[1]{%
    \begin{tikzpicture}[baseline={(current bounding box.center)}, rotate=45, scale=0.6]
        \fill[blue!30] (0,0) rectangle (0.8,0.8); 
        \draw (0,0) rectangle (0.8,0.8); 
    \end{tikzpicture}%
}

\newcommand{\re}{\text{Re}}
\begin{document}
\title{Non-Hermitian skin effect in arbitrary dimensions: non-Bloch band theory and classification}

\author{Yuncheng Xiong\textsuperscript{1}}
\altaffiliation{These authors contributed equally to this work.}
\author{Ze-Yu Xing\textsuperscript{1,2}}
\altaffiliation{These authors contributed equally to this work.}
\author{Haiping Hu\textsuperscript{1, 2}}\email{hhu@iphy.ac.cn}
\affiliation{\textsuperscript{1}Beijing National Laboratory for Condensed Matter Physics, Institute of Physics, Chinese Academy of Sciences, Beijing 100190, China}
\affiliation{\textsuperscript{2}School of Physical Sciences, University of Chinese Academy of Sciences, Beijing 100049, China}
\begin{abstract}
Non-Hermitian skin effect (NHSE) is a distinctive phenomenon in non-Hermitian systems, characterized by a significant accumulation of eigenstates at system boundaries. While well-understood in one dimension via non-Bloch band theory, unraveling the NHSE in higher dimensions faces formidable challenges due to the diversity of open boundary conditions or lattice geometries and inevitable numerical errors. Key issues, including higher-dimensional non-Bloch band theory, geometric dependency, spectral convergence and stability, and a complete classification of NHSE, remain elusive. In this work, we address these challenges by presenting a geometry-adaptive non-Bloch band theory in arbitrary dimensions, through the lens of spectral potential. Our formulation accurately determines the energy spectra, density of states, and generalized Brillouin zone for a given geometry in the thermodynamic limit (TDL), revealing their geometric dependencies. Furthermore, we systematically classify the NHSE into critical and non-reciprocal types using net winding numbers. In the critical case, we identify novel scale-free skin modes residing on the boundary. In the nonreciprocal case, the skin modes manifest in various forms, including normal or anomalous corner modes, boundary modes or scale-free modes. We reveal the non-convergence and instability of the non-Bloch spectra in the presence of scale-free modes and attribute it to the non-exchangeability of the zero-perturbation limit and the TDL. The instability drives the energy spectra towards the Amoeba spectra in the critical case. Our findings provide a unified non-Bloch band theory governing the energy spectra, density of states, and generalized Brillouin zone in the TDL, offering a comprehensive understanding of NHSE in arbitrary dimensions.
\end{abstract}
\maketitle
\section{Introduction}\label{sec1}
Non-Hermitian physics is a rapidly expanding field \cite{coll1,coll4,coll6,colladd3,nhreview,nhreview2,chlee2023b} that explores systems where the Hamiltonian is not Hermitian. Relevant across various classical wave systems in photonics \cite{photonic1,photonic2,op2,op5,op8,op17,op19,photonic3,photonic4,nhsee3,xuegeometry,xueinvariant}, acoustics \cite{ep2encircling3,epnexus,pg3,hep1,hep2,acoustic1,acoustic2}, mechanics \cite{mechanics1,mechanics2,mechanics3,mechanics4}, and electrical circuits \cite{nhsee2,zhangxd1,zhangxd2,zhangxd3}, non-Hermiticity also plays a crucial role in describing open quantum systems with nonconservative dynamics \cite{open1,open2,open3}. The breaking of Hermiticity leads to a myriad of intriguing phenomena absent in Hermitian systems. For instance, the energy spectra become highly sensitive to boundary conditions, with their eigenstates pushed towards system boundaries under open boundary conditions (OBC). This phenomenon, known as the non-Hermitian skin effect (NHSE) \cite{nhse1,nhse2,nhse3,nhse4,nhse5,nhse6,nhse7,nhse8,nhse9,nhse10,nhse11,nhsereview}, has been extensively studied both theoretically and experimentally \cite{nhsee1,nhsee2,nhsee3,nhsee4,nhsee5,nhsee6,zhangxd1,yiwei2022c,qiuchunyin2023a,qiuchunyin2024}, owing to its departure from Bloch band theory and its wide range of promising applications \cite{nhsea1,nhsefunnel,nhsea2,nhsea3,nhsea4,nhsea5}.

Unraveling the NHSE in one dimension (1D) led to the development of non-Bloch band theory \cite{nhse1,nhse4}, which accurately determines the asymptotic energy spectra under OBC in the thermodynamic limit (TDL) without suffering severe numerical errors \cite{nhse7}. By analytically continuing the Bloch Hamiltonian $H(k)\rightarrow H(\beta=e^{ik})$, the skin localization of eigenstates is described by the generalized Brillouin zone (GBZ), forming closed trajectories on the complex $\beta$-plane. The NHSE originates from intrinsic non-Hermitian point gaps \cite{nhse6,nhse5,pointtopo3} or spectral windings, as allowed by complex eigenenergies. Transitioning from the Brillouin zone to the GBZ captures the spectral collapse from loop-shaped Bloch bands \cite{hhp_knot} to open-arc-shaped non-Bloch bands on the complex-energy plane. An equivalent description of this collapse mechanism involves spectral potential formalism \cite{wz,hupotential}. Through the analogy with electrostatic Coulomb potential, both the spectral density of states (DOS) and the GBZ can be precisely reproduced from the potential landscape.

In higher dimensions, the diversity of OBC greatly complicates the NHSE and the spectral structures in the TDL. Different choices of OBC correspond to different lattice geometries \cite{huuniform}. The NHSE and non-Bloch spectra may highly depend on this geometric information \cite{fangchen,fangchen2,yangzhesen2023}. Besides the normal skin modes residing at systems' corners, higher-dimensional non-Hermitian systems may host edge skin modes \cite{fangchen,yzsedge}. To date, a comprehensive framework governing the non-Bloch bands and NHSE in various lattice geometries is still lacking. Key questions remain unsolved. For instance, given a Bloch Hamiltonian $H(\bm k)$, how can we determine the asymptotic spectra in the TDL under a given lattice geometry, and how are the skin modes manifested? Answering these questions is challenging, as the framework of 1D non-Bloch band theory cannot be directly extended to higher dimensions with infinite lattice geometries. A recent work proposed a higher-dimensional generalization of the GBZ condition via the Amoeba formulation \cite{wz}, which neglects geometric information and yields geometry-irrelevant non-Bloch spectra. Yet, insights from 1D NHSE suggest that boundary conditions or lattice geometry play a crucial role in the asymptotic spectral structure and skin modes. As will be demonstrated in this paper, in higher-dimensional non-Hermitian systems with clear boundaries, incorporating geometric information is essential to formulating the non-Bloch band theory.

The lack of a theoretical framework governing non-Bloch bands and NHSE in various lattice geometries has led to numerous unresolved questions and debates in the literature, as outlined below.

i)~\textit{Non-Bloch spectra.} Whether the asymptotic energy spectra and DOS in the TDL depend on geometry remains an open question under debates. In Ref. \cite{fangchen}, varying spectral densities associated with different lattice shapes are reported. However, numerical results in Ref. \cite{wz} suggest that systems under weak perturbations or with smooth boundaries should have the same DOS in the TDL. Refs. \cite{huuniform,wz,yzsconjecture} further conjecture that the energy spectra for any lattice geometry in the TDL should be given by the spectra from the Amoeba formulation. This debate seems difficult to settle since numerical errors arising from the non-normality of the Hamiltonians \cite{colladd3} are inevitable. Moreover, a key default assumption in previous research is the existence of non-Bloch spectra in the TDL. Yet, the spectral convergence must be scrutinized before developing the non-Bloch band theory.

ii)~\textit{GBZ.} The GBZ encodes the localization properties of skin modes. In 1D, it is analytically tractable and represented as closed loops in the 2D complex-$\beta$ plane \cite{nhse1,nhse4,nhse7}. In higher dimensions, finding the GBZ condition from boundary constraints is challenging. Using Ronkin's function, the GBZ is proposed \cite{wz} to be a $d$D object embedded in a $2d$D space for $d$D non-Hermitian systems without specifying the geometric information. Notably, Ref. \cite{jianghui} discovers dimensional transmutations, where the GBZ of certain 2D non-Hermitian models appears as 1D objects. Most recently, the GBZ with dimensions ranging from $d$ to $2d-1$ has been reported \cite{zhangkaigbz} and the geometry-relevance of the GBZ for certain models is revealed in Ref. \cite{yzsconjecture}. As the non-Bloch spectra and skin modes may be geometry-dependent, questions arise: Is the GBZ geometry-dependent? How can we determine the GBZ in a unified way that yields the asymptotic non-Bloch bands for a given geometry in the TDL?

iii)~\textit{Classification.} Various types of skin modes exist in high dimensions, e.g., corner- and edge-localized skin modes \cite{fangchen,yzsedge,zhangkaigbz}, and hybrid skin-topological modes \cite{llhhybrid,zhangxd1}. In certain cases, the appearance of skin modes is linked to the underlying geometric shapes, such as the geometry-dependent skin effect \cite{fangchen,yzsconjecture}. In 1D, a phenomenon known as critical NHSE features eigenstates whose localization varies with system size \cite{llhcritical}. To date, the universal properties of each type of skin mode and a criterion for classifying NHSE remain elusive. Additionally, it is still unclear whether higher-dimensional systems have counterparts to the 1D critical NHSE.

iv)~\textit{Stability.} Unlike Hermitian systems, whose energy spectra are stable against weak perturbations according to the Weyl perturbation theorem \cite{weyl}, non-Hermitian systems may exhibit extreme sensitivity to weak perturbations due to the non-orthogonality of eigenstates, e.g., in the presence of exceptional degeneracies \cite{colladd3,ep_review,hhp_ep1,hhp_ep2}. This spectral sensitivity can be harnessed to design functional sensors \cite{epsensor1,epsensor2,epsensor3,epsensor4}. Previous studies on non-Bloch band theory have rarely addressed the issue of spectral instability and its relation with the skin modes. In the presence of perturbations, such as random disorder, it remains to be seen whether the spectral properties differ significantly from those of the unperturbed system in the TDL. Moreover, what is the interconnection between spectral convergence and spectral instability?
\begin{figure}[!t]
\centering
\includegraphics[width=3.375 in]{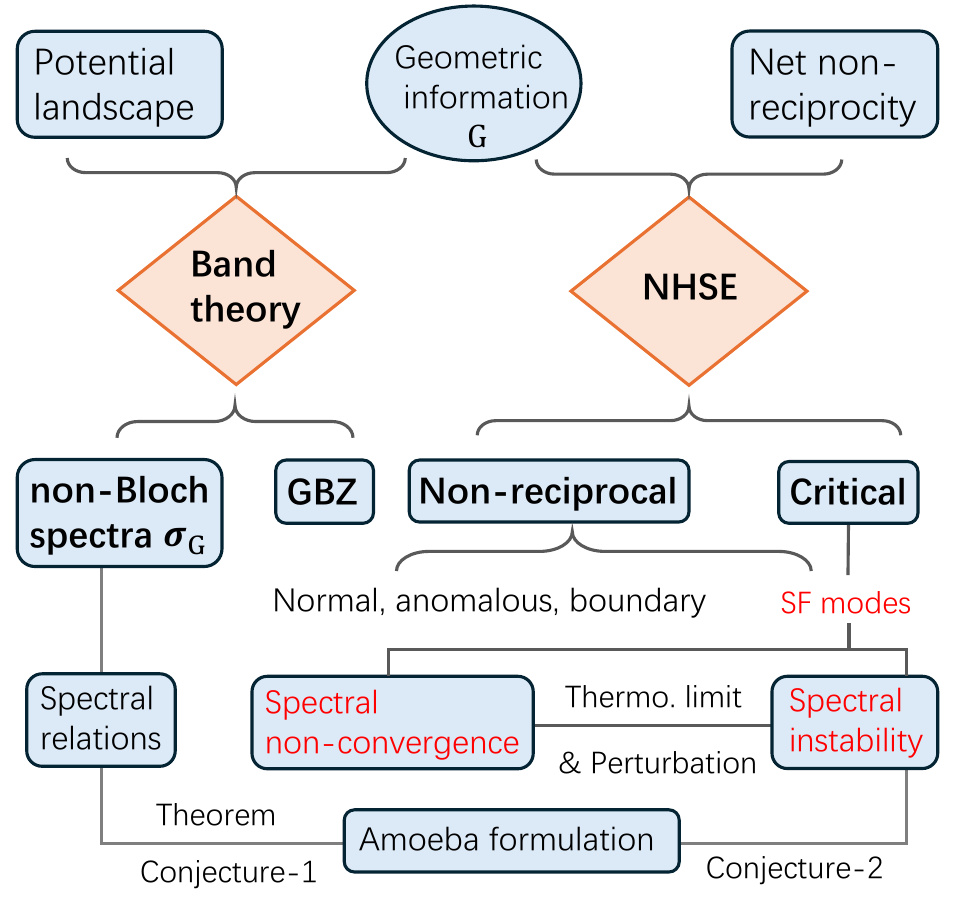}
\caption{Schematic of workflow. The non-Bloch band theory is developed from the potential landscape, incorporating geometric information. This approach yields the non-Bloch spectra, density of states, and generalized Brillouin zone (GBZ) in the thermodynamic limit, all of which are geometry-dependent. The net non-reciprocity and geometric information are combined to classify the NHSE into two types: critical and non-reciprocal. The former hosts scale-free (SF) skin modes, while the latter may host normal/anomalous corner modes, edge modes, or SF skin modes. The spectral instability under weak perturbations and the non-convergence in the thermodynamic limit are linked to the presence of scale-free modes. The relations between the non-Bloch spectra associated with different geometric shapes, the Amoeba spectra, and the perturbed spectra are established through a theorem and two conjectures.}\label{fig1}
\end{figure}

This article comprehensively addresses these questions by developing a non-Bloch band theory for arbitrary dimensions from the perspective of spectral potential. For regular lattice geometries, our formalism yields the asymptotic non-Bloch energy spectra, DOS, and GBZ in the TDL, demonstrating that all these quantities are geometry-dependent. Furthermore, we present a complete classification of NHSE through net winding numbers, identifying two types: critical and non-reciprocal. The critical case is the higher-dimensional counterpart of the 1D critical NHSE, hosting scale-free skin modes \cite{sfl1,sfl2,sfl3,sfl4}. In contrast, for the non-reciprocal case, the skin modes may manifest as normal/anomalous corner modes, edge modes, or scale-free modes. We demonstrate that in the presence of scale-free modes, the asymptotic energy spectra do not converge and exhibit instability when weak perturbations are introduced. We reveal that this spectral instability arises from the noncommutativity of the TDL and the zero-perturbation limit, and drives the perturbed energy spectra towards the Amoeba spectra in the critical case. Our work establishes a unified non-Bloch band theory that incorporates geometric information, laying the cornerstone for studying NHSE in all dimensions.

Our workflow is sketched in Fig. \ref{fig1}, and this article is organized as follows. Sections \ref{secii} to \ref{seciv} are dedicated to developing the non-Bloch band theory in arbitrary dimensions. Section \ref{secii} revisits the 1D NHSE and GBZ condition from two equivalent perspectives: the 1D non-Bloch band theory and the potential landscape. We highlight the main obstacles in formulating a higher-dimensional non-Bloch band theory and provide a guiding principle based on potential landscape. Section \ref{seciii} establishes the higher-dimensional non-Bloch band theory by extending the 1D potential landscape to higher dimensions hierarchically. Geometric shapes with different regular lattice cuts are related via basis transformations. We demonstrate that the spectral range and DOS are geometry-dependent, and our formulation agrees perfectly with numerical results. Section \ref{seciv} presents the GBZ condition for a given geometry from the potential landscape. Section \ref{secv} provides a pedagogical introduction to the Amoeba formulation. Through an analytically tractable model, we further benchmark our formalism and show its disparities with Amoeba. The relationships between different types of energy spectra and their geometric dependencies are established through a spectral theorem and the first conjecture.

We then shift gears to the NHSE in higher dimensions. Drawing insights from the 1D critical NHSE, Section \ref{secvi} establishes the criterion for the critical NHSE in higher dimensions through net winding numbers. We uncover that in the critical case, the skin modes exhibit scale-free localization, and the non-Bloch spectra in the TDL are not well-defined. Geometric factors like boundary ratios, may influence the spectral structures. Section \ref{secvii} addresses the non-reciprocal NHSE. Through concrete models, we illustrate various types of skin modes, e.g., normal/anomalous, boundary, and scale-free skin modes. The non-Bloch spectra are governed by our theory unless scale-free localization occurs.

The discussion on the non-Bloch bands and NHSE intertwines through spectral convergence and stability. Section \ref{secviii} addresses the spectral stability for both types of NHSE. We show that in the presence of scale-free modes, the zero-perturbation limit and the TDL do not commute. Thus, spectral non-convergence and instability are united through scale-free localization. Our numerics indicate that for the critical case, the energy spectra tend towards those of Amoeba formulation in the TDL, which constitutes our second conjecture. Section \ref{secix} offers a comprehensive classification of NHSE in all dimensions based on net winding numbers, summarized in Table \ref{table1}. Finally, in Section \ref{secx}, we conclude with the main findings, discuss their experimental relevance, and outline several open questions.

\section{Non-Bloch bands in 1D: a dual perspective} \label{secii}
\begin{figure}[!t]
\centering
\includegraphics[width=3.375 in]{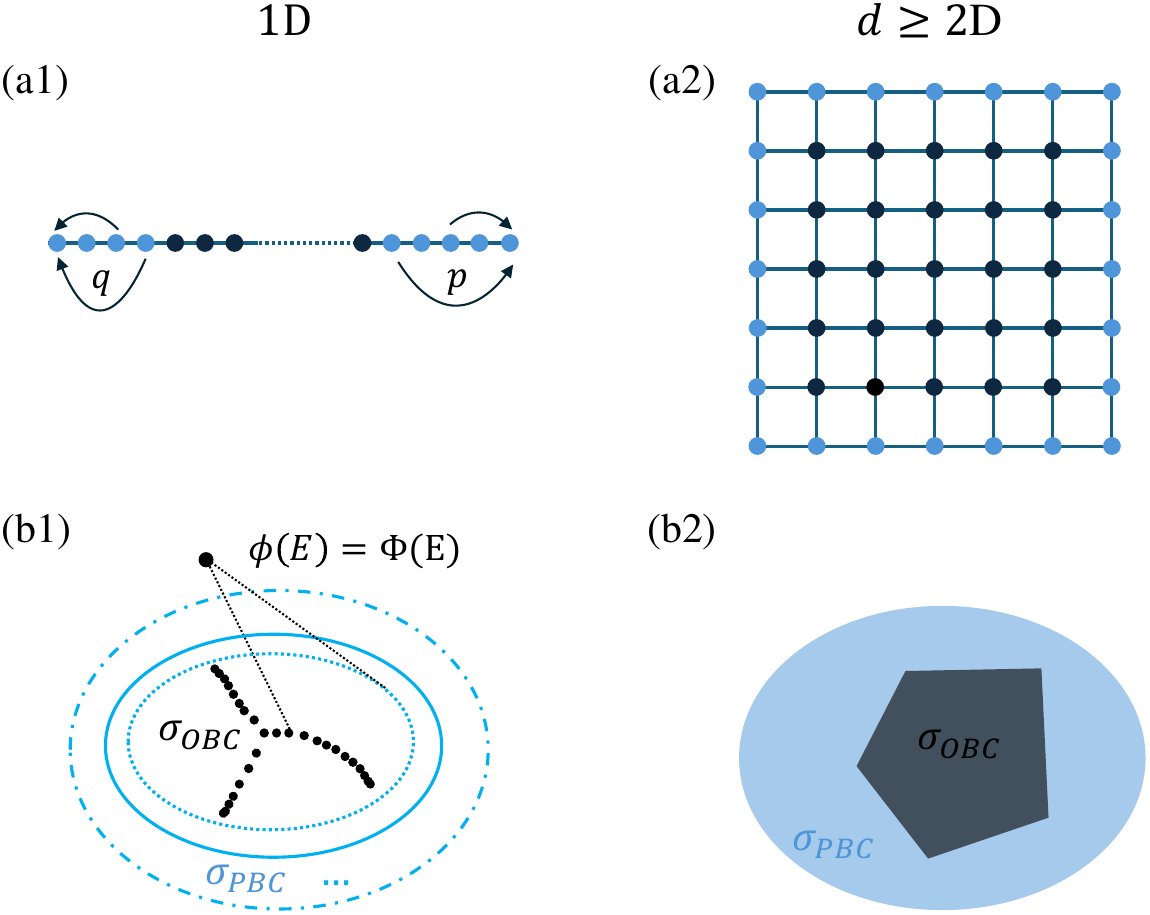}
\caption{Sketch of the challenges in formulating higher-dimensional non-Bloch band theory. (a1) A 1D lattice with the largest hopping range to the right/left as $p$ and $q$, respectively. The sites marked in blue are considered in the boundary equations. (a2) A higher-dimensional lattice with $O(L^{d-1})$ boundary sites. (b1) Schematics of the 1D potential theory. The OBC spectra/PBC spectra/deformed Bloch spectra are represented by black points/solid blue lines/dashed (dotted) lines, respectively. The OBC spectral potential $\phi(E)$ equals the electrostatic potentials generated by certain deformed spectra via Szeg\"o theorem. (b2) Sketch of the PBC/OBC spectra in higher dimensions, which typically occupy a finite area in the complex plane.}\label{fig2}
\end{figure}

We begin by reviewing how to obtain the non-Bloch spectra for generic 1D non-Hermitian lattice models. Non-Bloch spectra represent the continuum part of the energy spectra under OBC in the TDL. The 1D tight-binding Hamilotnian is given by
\begin{eqnarray}
H=\sum_{i,j=1}^N t_{i,j}c_{i}^{\dag} c_j,
\end{eqnarray}
where $c_i$ ($c_{i}^{\dag}$) denotes the annihilation (creation) operator on the $i$-th unit cell. If each unit cell has $s$ degrees of freedom (e.g., spin, sublattice, orbital), then $t_{ij}$ should be treated as an $s\times s$ matrix. With periodic boundary condition (PBC), the Fourier transformation takes the tight-binding Hamiltonian to its Bloch form: 
\begin{eqnarray}
H(k)=\sum_{m=-p}^q t_{m} e^{i mk},
\end{eqnarray}
where $t_{i-j}=t_{ij}$ and $p,q$ is the largest hopping range to the right/left, respectively, as depicted in Fig. \ref{fig2}(a1). $N$ is the total number of lattice sites. 

Under PBC, the eigenstates are extended Bloch waves. The energy spectra form closed loops in the complex plane. However, once OBC is applied, the NHSE occurs. The eigenstates become skin modes localized at two boundaries. The OBC energy bands form spectral arcs in the complex plane, as shown in Fig. \ref{fig2}(b1). To account for the presence of skin modes, the lattice momentum $k$ should be complexified as $k\rightarrow k+i \kappa$, where $\kappa$ denotes the inverse localization length of the skin modes. Let us set $\beta=e^{ik}$ , the Bloch Hamiltonian is analytically continued to $H(k)\rightarrow H(\beta)$ with characteristic polynomial (ChP)
\begin{eqnarray}
f(\beta,E)=\det|H(\beta)-E|.
\end{eqnarray}
The ChP has two complex variables. The non-Bloch spectra are given by the $E$-solutions of $f(\beta,E)=0$ when $\beta$ takes suitable values, which form the GBZ. Without loss of generality, we consider the single-band case ($s=1$) in the following.

There are two equivalent approaches to precisely determine the non-Bloch spectra and the GBZ: the 1D non-Bloch band theory and the potential landscape.

\subsection{Route 1: Non-Bloch band theory} \label{seciia}
We follow the standard procedure in Ref. \cite{nhse4}. Under OBC, the Hamiltonian is an $N\times N$ matrix. The eigenequation is $H|\psi\rangle=E|\psi\rangle$, with $|\psi\rangle=(\psi_1,\psi_2,...,\psi_N)^T$. The bulk components of the eigenfunction are
\begin{eqnarray}\label{eigeq}
\sum_{n=-p}^q t_n\psi_{j+n}=E\psi_j.
\end{eqnarray}
Taking an ansatz wavefunction $\psi_j\propto \beta^j$, we have $E=\sum_{n=-p}^q t_n \beta^n$. For a given $E$, there are $p+q$ (complex) solutions of $\beta$, giving rise to $p+q$ degenerate bulk wavefunctions $\psi^{(l)}(l=1,\cdots,p+q)$. These solutions can be sorted according to their moduli: $|\beta_1(E)|\leq|\beta_2(E)|\leq\cdots \leq|\beta_{p+q}(E)|$. In addition to the bulk equations, there are $p+q$ boundary equations subjected to the OBC:
\begin{eqnarray}\label{boundaryeq}
&&\sum_{n=i}^q t_n \psi_{j+n} =E\psi_j,	\quad i=-p+1,-p+2,...,0, \nonumber\\
&&\sum_{n=-p}^i t_n \psi_{j+n}=E\psi_j, \quad i=0,1,...,q-1.
\end{eqnarray}
To construct the true wavefunctions satisfing these boundary equations, the degenerate bulk eigenfunctions should be superimposed properly, i.e., $\psi_j=\sum_{l=1}^{p+q} c_l \psi_j^{(l)}=\sum_{l=1}^{p+q} c_l \beta_l^j$. The boundary equations are then recast in a matrix form:
\begin{eqnarray}\label{boundaryeq}
M (c_1,c_2,...,c_{p+q})^T=0.
\end{eqnarray}
Here $M$ is a $(p+q)\times(p+q)$ matrix. Nontrivial solutions of the boundary equations requires $\det M=0$. For large system size $N$, the two dominant terms in the determinant are \cite{nhse4}
\begin{align}
	(\cdot)_1\big(\beta_{p+q}\beta_{p+q-1}\cdots\beta_{p+2}\beta_{p+1}\big)^N \nonumber\\
	+(\cdot)_2\big(\beta_{p+q}\beta_{p+q-1}\cdots\beta_{p+2}\beta_{p}\big)^N,
\end{align}
where we have omitted the $N$-irrelevant coefficients denoted by $(\cdot)_{1,2}$. If $|\beta_{p}|<|\beta_{p+1}|$, the first term is exponentially larger than the second one as $N\to\infty$. The vanishing of the first coefficient $(\cdot)_1$ produces only finite number of eigenenergies due to its $N$-independence. Thus, for the continuum non-Bloch bands, we must have the following relation:
\begin{eqnarray}\label{gbz1d}
|\beta_{p}(E)|=|\beta_{p+1}(E)|,
\end{eqnarray}
which is the GBZ condition of 1D non-Bloch bands.

\subsection{Route 2: Potential lanscape in 1D} \label{seciib}
Instead of directly solving the eigenequation (\ref{eigeq}), the logarithmic potential theory \cite{wz,hupotential} offers an elegant formulation of the non-Bloch spectra, DOS, and GBZ for 1D non-Hermitian systems. Let us denote the eigenenergies of the OBC Hamiltonian as $E_i \ (i=1,2,...,N)$. The spectral DOS in the TDL is
\begin{align}
\rho(E)=\frac{1}{N}\sum_{i=1}^N \delta_{E,E_i},
\end{align}
with Kronecker delta function $\delta_{E,E_i}=1$ if $E=E_i$. By treating each eigenenergy as a particle with charge $1/N$ in the complex plane, the electrostatic potential felt at position $E \in \mathbb{C}$ is [See Fig. \ref{fig2}(b1)]
\begin{eqnarray}\label{def_pot1}
\phi(E)=\frac{1}{N}\sum_{i=1}^N\log|E_i-E|.
\end{eqnarray}
In the TDL, the DOS is related to the electrostatic potential via the Poisson equation:
\begin{eqnarray}\label{po_eq}
\rho(E)=\frac{1}{2\pi}\nabla^2_E \phi(E).
\end{eqnarray}

To obtain the non-Bloch spectra, the key observation is that in the TDL, the potential function $\phi(E)$ takes the integral form:
\begin{eqnarray}
\lim_{N\rightarrow\infty}\phi(E)=\int^{2\pi}_0 \frac{dk}{2\pi} \log|\det[H(e^{ik+\mu})-E]|, \label{1dpotential} 
\end{eqnarray}
according to Szeg\"o limit theorem \cite{szego1,szego2}. Here, the parameter $\mu$ should be properly chosen such that $|\beta_p(E)| < e^\mu < |\beta_{p+1}(E)|$. Notably, this condition is not satisfied if $E$ is located on the spectral arcs due to $|\beta_p(E)| = |\beta_{p+1}(E)|$. Setting aside this detail for now and performing the integral in Eq. (\ref{1dpotential}), one can obtain the local form of the electrostatic potential \cite{hupotential}:
\begin{align} 
\lim_{N\rightarrow\infty}\phi(E)=\sum_{j=p+1}^{p+q}\log|\beta_j(E)|+\log|t_q|, \label{1dlocal}
\end{align}
where $\beta_j$ are the zeros of ChP sorted as $|\beta_1|\leq|\beta_2|\leq\cdots \leq|\beta_{p+q}|$. Thus the potential $\phi(E)$ has contributions from the $q$ roots of the largest moduli. Analysis of the harmonicity of the spectral potential in Eq. (\ref{1dlocal}) leads to the GBZ condition $|\beta_p(E)|=|\beta_{p+1}(E)|$ \cite{hupotential}. We note that while Szeg\"o’s limit theorem fails when $E$ is inside the non-Bloch spectra in Eq. (\ref{1dpotential}), the local form in Eq. (\ref{1dlocal}) works for any $E$ in the complex plane. For a given $E$, we only need to find the roots of the ChP $f(\beta,E)=0$ to obtain the potential function $\Phi(E)$. Through Eq. (\ref{po_eq}), the spectral range and DOS of the non-Bloch bands are determined.

A physical interpretation of the potential theory is as follows. We deform the Bloch spectra under PBC, i.e., $H(k\rightarrow k-i\mu)$, and consider the electrostatic potential generated by the deformed loops. Figure \ref{fig2}(b1) sketches the PBC spectra and several deformed spectra. Szeg\"o theorem in Eq. (\ref{1dpotential}) states that, the electrostatic potential generated by the non-Bloch spectra under OBC in the TDL equals the electrostatic potential generated by certain deformed spectra. By scanning all possible deformation parameters $\mu \in (-\infty,+\infty)$, it can be rigorously proven [See Appendix \ref{appendixa} for the proof] that
\begin{eqnarray}\label{1d_potential}
&&\lim_{N\rightarrow\infty}\phi(E)\notag\\
&&=\min_\mu \int^{2\pi}_0 \frac{d k}{2\pi} \log|\det[H(e^{i k+\mu})-E]|, \forall E. \label{1dmin}
\end{eqnarray}
This means that the local potential at $E$ is the minimum among all possible deformed spectral loops. Thus we propose the following principle of non-Bloch bands from the perspective of potential landscape.\\
\textbf{Guiding principle}:
Among all possible spectral deformations, the one with the minimum spectral potential corresponds to the potential generated by the non-Bloch bands.

\subsection{Challenges in extending to higher dimensions} \label{seciic}
One primary challenge in formulating higher-dimensional non-Bloch band theory is the complexity of OBC or the diversity of geometric shapes. Back to the two approaches above, both accurately determine the non-Bloch bands in the TDL and the GBZ in 1D. However, extending them to higher dimensions poses significant challenges. In 1D, the ChP has a finite number ($p+q$) of $\beta$-solutions. In the first route, the skin modes are constructed from the superposition of these solutions subjected to the boundary equations. For a $d$D Bloch Hamiltonian $H(k_1,k_2,\ldots,k_d)$, let us analytically continue the lattice momentum and consider the Hamiltonian $H(\beta_1,\beta_2,\ldots,\beta_d)$ with the ChP
\begin{eqnarray} 
f(\beta_1,\beta_2,\ldots,\beta_d,E)=\det[H(\beta_1,\beta_2,\ldots,\beta_d)-E].
\end{eqnarray} 
For a given $E$, there exist infinite solutions of $(\beta_1,\beta_2,\ldots,\beta_d)$. Moreover, OBC yields a finite number of boundary equations in 1D as sketched in Fig. \ref{fig2}(a1). The GBZ condition is derived from these boundary constraints. However, in $d\geq 2$D, the number of boundary equations is of $O(L^{d-1})$ order ($L$ is the linear length of the lattice) [See Fig. \ref{fig2}(a2)], which tends to infinity in the TDL. Thus, it is impossible to simultaneously solve these boundary equations and determine the GBZ condition.

For the second route, the potential theory relies on Szeg\"o limit theorem. In higher dimensions, the theorem reads \cite{szego1,szego2,widom1980,doktorskii1984}: 
\begin{eqnarray}
\lim_{N\rightarrow\infty}\phi(E)=\int \frac{d^d\bm k}{(2\pi)^d} \log|f(e^{i{\bm k}+{\bm\mu}},E)|, \label{szegodd} 
\end{eqnarray}
with $\bm k=(k_1,k_2,\ldots,k_d)$ and $\bm\mu=(\mu_1,\mu_2,\ldots,\mu_d)$. Here, $\bm\mu$ should be properly chosen such that the spectral windings along all $d$ directions vanish, a detail not elaborated here. Similar to 1D, if $E$ is inside the spectral region, Szeg\"o’s theorem fails. In 1D with arc-shaped spectra, Szeg\"o’s theorem can be analytically extended to the entire complex plane, i.e., via the neat local form in Eq. (\ref{1dlocal}). However, the OBC spectra of higher-dimensional systems typically occupy a finite area in the complex plane, as shown in Fig. \ref{fig2}(b2). Analytic continuation of Szeg\"o’s theorem to the whole complex plane becomes indefinite. We will discuss a tactful attempt through Amoeba formulation in Section \ref{secva}, although the geometric information is lost.

\section{Non-Bloch band theory in higher dimensions} \label{seciii}
In this section, we formulate the non-Bloch band theory hierarchically from the perspective of the potential landscape. The geometric information of lattice cuts is incorporated through basis transformations. This approach allows us to derive the potential function $\Phi(E)$ for generic $d$D non-Hermitian systems with regular geometric shapes and obtain the DOS in the TDL via the Poisson equation in a unified manner. Our formulation is then compared to numerical results, showing perfect agreement. It is important to note that for a given geometric shape, the non-Bloch spectra as the asymptotic limit of the OBC energy spectra may not converge, and therefore may not be well-defined. We will delve into the issues of spectral convergence and stability and address their conditions in Section \ref{secviii}.

\subsection{Geometric information} \label{seciiia}
\begin{figure}
	\includegraphics[width=3.375in]{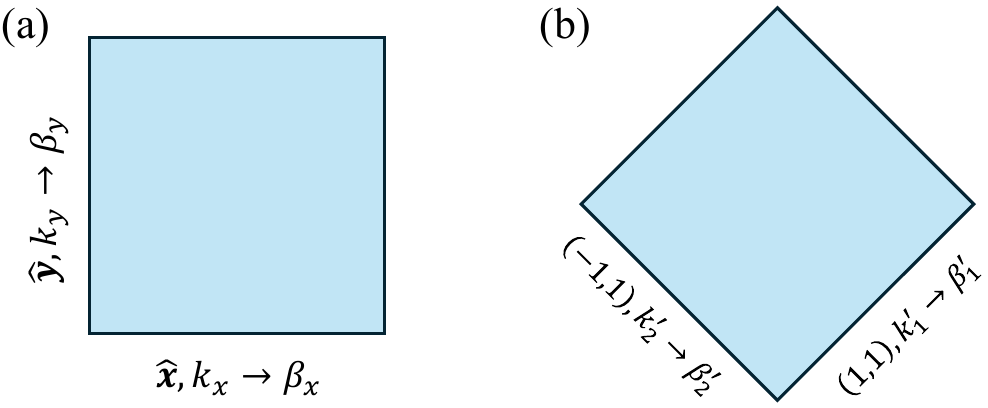}
	\caption{\label{fig3}Sketch of basis transformations that relate different geometric shapes. For the (a) square and (b) rhombus shapes, the transformation is $k^\prime_1 = (k_x + k_y)/2$ and $k^\prime_2 = (-k_x + k_y)/2$. The non-Bloch spectra for a regular shape are obtained via analytical continuation along its lattice-cut directions.}
\end{figure}

The geometric information refers to the shape of the lattice considered, e.g., the orientations of lattice cuts. Unlike the 1D case where OBC has a definite meaning, geometric shapes in higher dimensions are diverse. For instance, in 2D, cutting the lattice along $x$ and $y$ axes or along the diagonal directions respectively yields square and rhombus geometries, as depicted in Fig. \ref{fig3}. Previous model studies \cite{fangchen, yzsedge} have shown that geometric shape may influence the structure of the non-Bloch spectra. Therefore, a proper consideration of geometric information is crucial for formulating the non-Bloch band theory. The primary issue is how to handle different geometric shapes in a unified way.

To simplify the problem, we mainly focus on regular geometric shapes (e.g., parallelograms in 2D and parallelepipeds in 3D), which have $d$ independent cut directions on a $d$D lattice. We label the cut along direction $\hat{v}$ as cut-$\hat{v}$. Geometries with smooth boundaries (e.g., disk/ball in 2D/3D) or irregular shapes (e.g., irregular pentagon) will be addressed in Section \ref{secv}. Given a Bloch Hamiltonian $H(\bm k)$, different regular geometries are related by basis transformations \cite{huuniform}. Taking a 2D lattice as an example, Fig. \ref{fig3} illustrates how to deal with square and rhombus shapes. When examining the skin effect on the square geometry in Fig. \ref{fig3}(a), the procedure begins by complexifying lattice momenta $k_x,k_y$ along the two cuts. That is, we perform analytical continuation $H(k_x,k_y)\rightarrow H(\beta_x,\beta_y)$. The localization lengths along the cut-$\hat{x}$ and cut-$\hat{y}$ directions are encoded in $(\log|\beta_x|, \log|\beta_y|)$. For the rhombus geometry, the skin localization is along the $(1,1)$ and $(-1,1)$ directions. Thus, analytical continuation should be performed for the lattice momenta along these two directions
 \begin{eqnarray}
H(k_1^\prime , k_2^\prime )\rightarrow H(\beta_1^{\prime},\beta_2^{\prime}), 
\end{eqnarray}
with $k^\prime_1 = (k_x + k_y)/2$ and $k^\prime_2 = (-k_x + k_y)/2$. The basis transformation gives rise to the transformation relations of the complexified momenta: $\beta_1 = \beta_1^\prime \beta_2^{\prime -1}$ and $\beta_2 = \beta_1^\prime \beta_2^\prime$. The basis transformation enables us to deal with any regular geometries via suitable analytical continuation.

\subsection{Spectral potentials in higher dimensions} \label{seciiib}
We consider a general $d$D Bloch Hamiltonian $H(\bm k) = H(k_1, k_2, \ldots, k_d)$. The lattice geometry is specified by the choice of the set of lattice momenta ${k_1, k_2, \ldots, k_d}$. Our aim is to obtain the non-Bloch spectra associated with this geometric shape in the TDL. To this end, we perform the analytical continuation $H(k_1, k_2, \ldots, k_d) \rightarrow H(\beta_1, \beta_2, \ldots, \beta_d)$ with $\beta_j = e^{ik_j}$ for $j = 1, 2, \ldots, d$. Starting from the 2D case, we formulate the spectral potential and then extend it to higher dimensions hierarchically.

\subsubsection{\textit{$d=2$D.}}
\begin{figure}[!t]
\centering
\includegraphics[width=3.375 in]{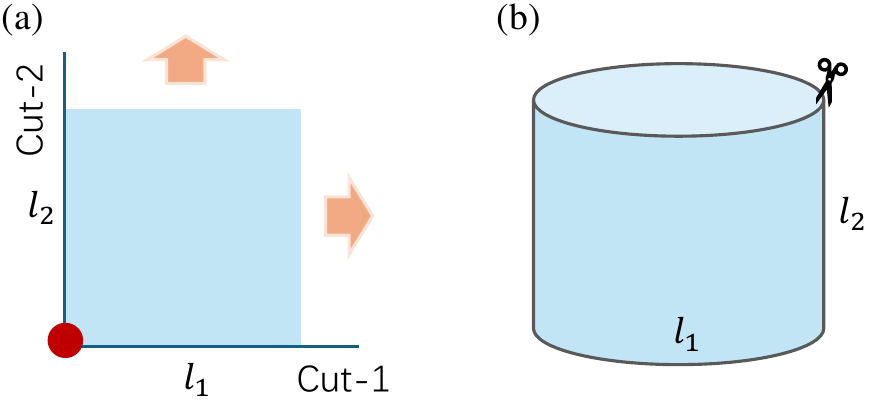}
\caption{Schematics of the formulation of the spectral potential. (a) A 2D lattice with two cut directions. When the system is large ($1\ll l_1, l_2<\infty$), further increasing the size (orange arrows) does not affect the skin modes (red dot) and the eigenspectra have stabilized, as required by the convergence condition. (b) Cylindrical geometry with PBC along the cut-1 direction and OBC along the cut-2 direction. The Hamiltonian is labeled as $\tilde{H}_{k_1}$.}\label{fig4}
\end{figure}

The 2D non-Hermitian Hamiltonian, after analytical continuation, takes the generic form:
\begin{eqnarray}
H(\beta_1,\beta_2)=\sum_{n_1=-p_1}^{q_1} \sum_{n_2=-p_2}^{q_2} f_{n_1,n_2}\beta_1^{n_1}\beta_2^{n_2},
\end{eqnarray}
where $p_1, q_1$ and $p_2, q_2$ denote the largest hopping ranges along cut-1 and cut-2 directions, respectively. For the square shape, $p_1, q_1,p_2,q_2$ refer to ranges to the right, left, upward, and downward. We take a finite 2D lattice with a total number of sites $N$. The lengths along the two cuts are denoted as $l_1$ and $l_2$, as shown in Fig. \ref{fig4}(a). Under OBC, the system's eigenenergies are labeled as $E_1, E_2, \ldots, E_{N}$, and their spectral potential is given by Eq. (\ref{def_pot1}). We temporarily neglect the influence of the boundary ratio $l_1/l_2$ on the non-Bloch spectra and skin modes. This is because, for normal skin modes, the localization lengths along the lattice cuts are definite and independent of the system size. However, this condition is violated when scale-free localization occurs. A detailed discussion of the boundary-ratio issue will be provided in Section \ref{secvid}.

A key observation is that, when $l_1$ and $l_2$ are sufficiently large, the eigenspectra and eigenstates should become \textit{very close} to the non-Bloch spectra and skin modes in the TDL. This indicates the \textit{convergence} of the non-Bloch spectra in the TDL. We proceed by scrutinizing the cylindrical geometry depicted in Fig. \ref{fig4}(b), where the cut-1 direction has PBC while the cut-2 direction has OBC. In this case, the lattice momentum $k_1$ remains a good quantum number. Formally, let us perform analytical continuation for the Bloch Hamiltonian $H(k_1,k_2)$ solely along the cut-2 direction and rewrite the Hamiltonian as 
\begin{eqnarray} H_{k_1}(\beta_2)=\sum_{n_2=-p_2}^{q_2}f_{n_2}^{(2)}(k_1)\beta_2^{n_2}, 
\end{eqnarray}
with $f_{n_2}^{(2)}(k_1)=\sum_{n_1=-p_1}^{q_1} f_{n_1,n_2}e^{in_1 k_1}$. Here $k_1$ is treated as a parameter. For each $k_1$-slice, utilizing the potential formula developed in Section \ref{seciib}, we obtain its spectral potential: 
\begin{eqnarray}\label{potential_quasi1d} 
\Phi_{k_1}(E)=\sum_{n=p_2+1}^{p_2+q_2}\log|\beta_{2,n}(k_1,E)|+\log|f_{q_2}^{(2)}(k_1)|. 
\end{eqnarray}
Here $\beta_{2,n}$ is the $n$-th zero of the ChP $\det[H_{k_1}(\beta_2)-E]=0$ with respect to $\beta_2$ ($k_1$ is treated as a parameter) sorted by their moduli $|\beta_{2,1}|\leq|\beta_{2,2}|\leq\cdots\leq|\beta_{2,p_2+q_2}|$. Obviously, $\Phi_{k_1}(E)$ represents the spectral potential contributed from the $k_1$-slice under the cylindrical geometry when $l_2\rightarrow\infty$.

From another perspective, this cylindrical system can be viewed as a quasi-1D lattice system of $l_2$ bands under PBC along the cut-1 direction. The Hamiltonian is an $l_2 \times l_2$ matrix parameterized by the lattice momentum $k_1$, denoted as $\tilde{H}_{k_1}$. Now, let's examine the scenario where the cut-1 direction is also open and apply the 1D potential theory in Section \ref{seciib} to this multi-band system. According to Eq. (\ref{1dmin}), the spectral potential after opening the cut-1 direction is given by: 
\begin{eqnarray}\label{potential_obc1}
\tilde{\Phi}_1(E) &=& \min_{\mu_1} \int_0^{2\pi} \frac{d k_1}{2\pi} \log \left| \text{det}[\tilde{H}_{k_1 -i \mu_1} - E] \right| \notag \\
&=& \min_{\mu_1} \int_0^{2\pi} \frac{d k_1}{2\pi} \Phi_{k_1-i\mu_1}(E)\notag\\
&=& \min_{\mu_1} \int_0^{2\pi} \frac{d k_1}{2\pi}\Big[ \sum_{n=p_2+1}^{p_2+q_2} \log |\beta_{2,n}(k_1-i\mu_1,E)|\notag\\
&& + \log |f_{q_2}^{(2)}(k_1-i\mu_1)|\Big].
\end{eqnarray}
Note that the integrand $\log|\det[…]|$ in the first line represents the spectral potential coming from the slice parameterized by $k_1-i\mu_1$. And in the second and third lines, we have substituted it with Eq. (\ref{potential_quasi1d}).

In a similar vein, we can consider another cylindrical geometry with OBC along the cut-1 direction and PBC along the cut-2 direction. To this end, we analytically continue the Bloch Hamiltonian $H(k_1,k_2)$ solely along the cut-1 direction and rewrite the Hamiltonian as $H_{k_2}(\beta_1)=\sum_{n_1=-p_1}^{q_1}f_{n_1}^{(1)}(k_2)\beta_1^{n_1}$, with $f_{n_1}^{(1)}(k_2)=\sum_{n_2=-p_2}^{q_2} f_{n_1,n_2}e^{in_2k_2}$. Following the same procedure as Eqs. (\ref{potential_quasi1d})(\ref{potential_obc1}), we thus obtain the spectral potential after opening the cut-2 direction:
\begin{align}
	\tilde{\Phi}_2(E)=\min_{\mu_2} \int_0^{2\pi}\frac{d k_2}{2\pi} \Big[&\sum_{n=p_1+1}^{p_1+q_1}\log|\beta_{1,n}(k_2-i\mu_2,E)| \nonumber\\
	&+\log|f_{q_1}^{(1)}(k_2-i\mu_2)|\Big].
	\label{eq:ourpotential_y}
\end{align}
There is no guarantee that the two potentials $\tilde{\Phi}_1(E)$, $\tilde{\Phi}_2(E)$ should be equal to each other. Based on our guiding principle in Section \ref{seciib}, we should choose the smaller one of the two potentials. Thus, we arrive at the final expression for the spectral potential:
\begin{eqnarray} \label{2d_pot}
\Phi(E) = \min\{\tilde{\Phi}_1(E),\tilde{\Phi}_2(E)\},
\end{eqnarray}
which corresponds to the lattice geometry with OBC along both the cut-1 and cut-2 directions. 

\subsubsection{\textit{$d>2$D}}
The 2D formulation extends to higher dimensions straightforwardly. In $d$D, the analytically continued Hamiltonian takes the general form:
\begin{eqnarray}
&&H(\beta_1, \beta_2, \ldots, \beta_d)\notag\\
&& = \sum_{n_j=-p_j;j=1,2,\ldots,d}^{q_j} f_{n_1, n_2, \ldots, n_d} \beta_1^{n_1} \beta_2^{n_2} \ldots \beta_d^{n_d}.
\end{eqnarray}
For systems with large size, the eigenspectra and eigenstates typically (except when scale-free localization occurs) stabilize and are very close to the non-Bloch spectra and skin modes in the TDL. Drawing insights from the 2D case, for a $d$D system, we consider a scenario where the cut-1 direction has PBC, but the remaining $(d-1)$ lattice cuts have OBC, with lengths $l_2, l_3, \ldots, l_d$. We assume $1 \ll l_2, l_3, \ldots, l_d < \infty$. In this case, the lattice momentum $k_1$ is a good quantum number. We perform analytical continuation along all directions except the cut-1. The $k_1$-slice Hamiltonian can be rewritten as
\begin{eqnarray} \label{dd_potential_quasi1d}
H_{k_1}(\beta_2, \ldots, \beta_d) = \sum_{n_j=-p_j; j=2,\ldots,d}^{q_j} f_{n_2, \ldots, n_d}^{(2, \ldots, d)}(k_1) \beta_2^{n_2} \ldots \beta_d^{n_d}.\notag\\ 
\end{eqnarray}
with $f_{n_2, \ldots, n_d}^{(2, \ldots, d)}(k_1) = \sum_{n_1 = -p_1}^{q_1} f_{n_1, n_2, \ldots, n_d} e^{i n_1 k_1}$. For each $k_1$-slice, we utilize the potential formulation of $(d-1)$D in a hierarchical way (e.g., Eq. (\ref{2d_pot}) in 2D. For consistency in notation, we still label it as $\Phi_{k_1}(E)$.

Then we change the perspective. The system can be regarded as a quasi-1D lattice of $(l_2 \ldots l_d)$ bands under PBC along the cut-1 direction. We label the Hamiltonian as $\tilde{H}_{k_1}$, which is an $(l_2 \ldots l_d) \times (l_2 \ldots l_d)$ matrix parameterized by the lattice momentum $k_1$. Once the cut-1 direction takes OBC, the 1D potential theory developed in Section \ref{seciib} can be applied to this multi-band system. From Eq. (\ref{1dmin}), the spectral potential after opening the cut-1 direction is: 
\begin{eqnarray} \label{dd_potential_obc1}
\tilde{\Phi}_1(E) &=& \min_{\mu_1} \int_0^{2\pi} \frac{d k_1}{2\pi} \log \left| \text{det}[\tilde{H}_{k_1 -i \mu_1} - E] \right| \notag \\
&=& \min_{\mu_1} \int_0^{2\pi} \frac{d k_1}{2\pi} \Phi_{\theta_1-i\mu_1}(E).
\end{eqnarray}
Similarly, we can deal with all the other $(d-1)$ cylindrical geometries, where the $j$-th one has PBC solely along the cut-$j$ $(j=2,\ldots,d)$ direction and OBC along all other directions. Following the same procedure as in Eqs. (\ref{dd_potential_quasi1d})(\ref{dd_potential_obc1}), the resulting spectral potential is denoted as $\tilde{\Phi}_j(E)$. According to the guiding principle in Section \ref{seciib}, we take the smallest of all these $d$ potentials and arrive at the final expression:
\begin{eqnarray}\label{dd_pot}
\Phi(E) = \min\{\tilde{\Phi}_1(E), \tilde{\Phi}_2(E), \ldots, \tilde{\Phi}_d(E)\}.
\end{eqnarray}
This spectral potential corresponds to the lattice geometry with $d$ cuts specified by the lattice momenta $k_1, k_2, \ldots, k_d$ in the TDL.

\subsubsection{\textit{Remarks}}
Firstly, our formulation of the spectral potential starts from Szeg\"o theorem in 1D. It is exact and well-defined across the entire complex plane in its local form, as shown in Eq. (\ref{1dlocal}). The analysis is then conducted hierarchically to derive the spectral potential for $d\geq2$D. No prior assumptions are made about the higher-dimensional Szeg\"o theorem, considering it cannot be applied to the spectral region, as mentioned in Section \ref{seciic}. This differs from the Amoeba formulation \cite{wz}, where certain minimization of Ronkin’s function is assumed. In fact, it can be rigorously proven that our potential function in Eqs. (\ref{2d_pot}) and (\ref{dd_pot}) is always no greater than the spectral potential in the Amoeba formulation, which will be addressed in Section \ref{secv}. Additionally, we will further discuss the relationship between the Amoeba spectra and the genuine non-Bloch spectra in that section.

Secondly, we note that the spectral potential depends on the specific choice of geometry. Through the Poisson equation in Eq. (\ref{po_eq}), the DOS of the system under this geometry in the TDL is obtained. Different lattice cuts may yield different spectral potentials. Consequently, different geometries can exhibit distinct spectral ranges and DOS, which will be confirmed in various models in the following sections. This sharply contrasts with the spectra obtained from the Amoeba formulation, which is geometry-independent. From now on, we denote 
\begin{eqnarray}
\sigma_G:~~\text{non-Bloch spectra under geometry $G$}
\end{eqnarray}
in the TDL. For regular geometries, the shape is specified by the lattice-cut orientations. Note that $\sigma_G$ contains information about both the spectral range and the DOS. In some cases, we may only be concerned with the spectral range in a set-theoretical sense and treat $\sigma_G$ as a set.

Thirdly, the only assumption in our formulation is the spectral convergence, i.e, the stabilization of systems' eigenspectra and eigenstates for large system sizes. Our theory then predicts that they would converge to the non-Bloch spectra and skin modes in the TDL. However, this assumption is not valid when the system exhibits scale-free localization, where the localization length of skin mode scales with system size. This represents a critical case, and we will illustrate that for such systems, the eigenspectra highly depend on system size and boundary ratios and exhibit instability in the presence of perturbations. Thus, the non-Bloch spectra in the TDL are not well-defined solely with the information on the lattice-cut directions. We will delve into these scenarios extensively in Sections \ref{secvi} and \ref{secvii} and elucidate the consequences of scale-free localization.

\subsection{An illustrating example}\label{seciiic}
\begin{figure}
\includegraphics[width=3.375in]{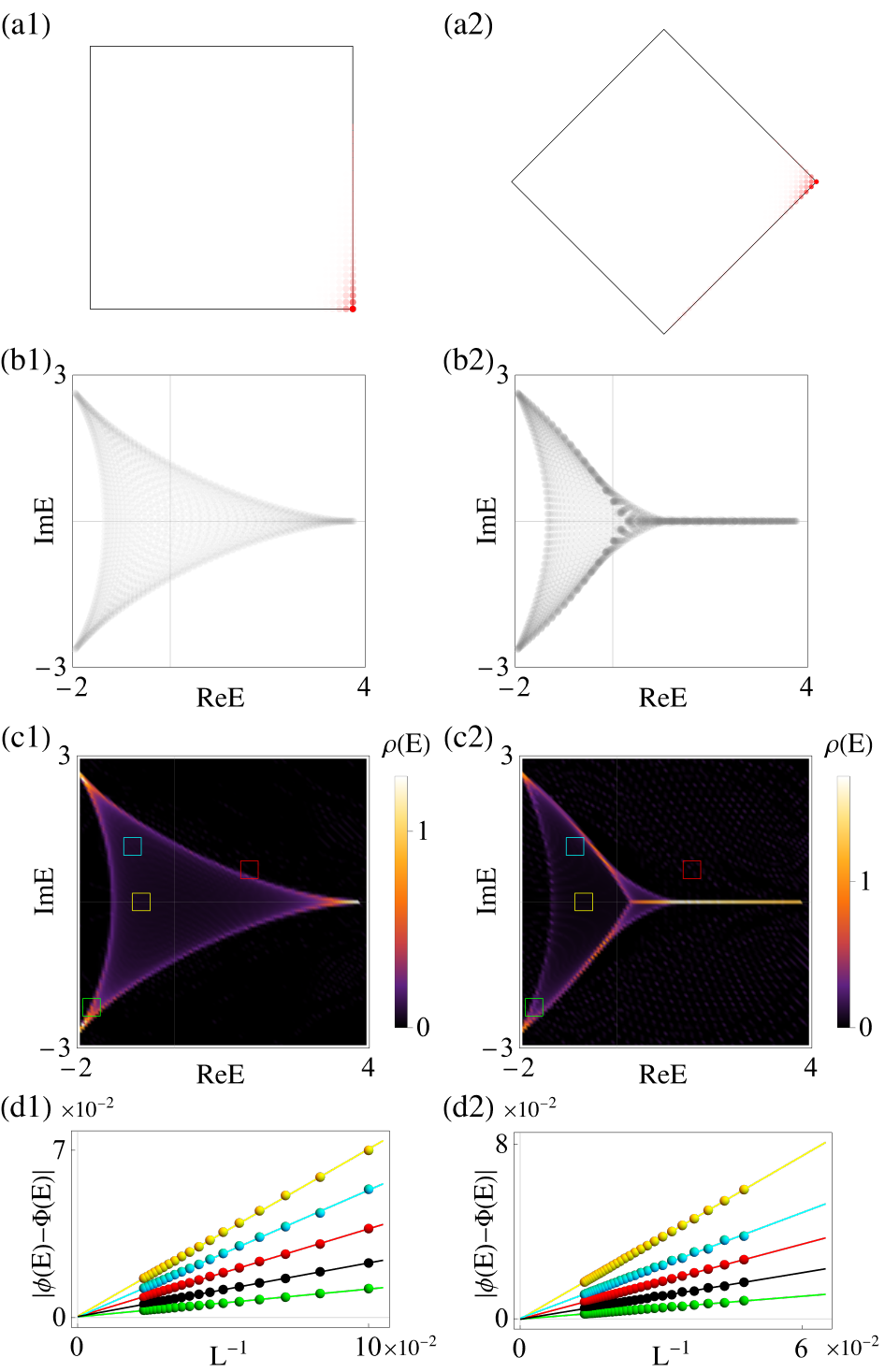}
\caption{\label{fig5}Geometry-dependence of the non-Bloch spectra in the TDL. (a1, a2) Spatial distributions of all eigenstates under the square and rhombus geometries, respectively. (b1, b2) Energy spectra for the model (\ref{cse_model}) obtained from exact diagonalization. The lattice sites are $N=1600$ (square) and $N=1861$ (rhombus), respectively. (c1, c2) Spectral density of states (DOS) obtained by solving the Poisson equation using our potential function $\Phi(E)$ in Eq. (\ref{2d_pot}). The two geometries have different spectral ranges and DOS. (d) Finite-size analysis of the average absolute deviations of the spectral potential $|\phi(E)-\Phi(E)|$ for the boxed regions in (c1) and (c2). $L$ is the boundary length. $49$ energy points are chosen for each box. The black dots represent averages over the entire complex plane. $\phi(E)$ is obtained from exact diagonalization.}
\end{figure}

We illustrate our potential formulation of non-Bloch bands with the following model Hamiltonian:
\begin{eqnarray}\label{cse_model}
H(\beta_x,\beta_y)=2\beta_x^{-1}+\frac{1}{2}\beta_x\beta_y^{-1}+\frac{3}{2}\beta_y+\frac{9}{10}\beta_x^{-1}\beta_y.
\end{eqnarray}
Let us present the numerical results first. We consider two distinct lattice geometries, namely square and rhombus, and plot the spatial distributions of the system's eigenstates in Figs. \ref{fig5}(a1)(a2). For both shapes, we observe well-localized skin modes at the corners. In Figs. \ref{fig5}(b1)(b2), we show the eigenspectra obtained from exact diagonalization for both shapes. Notably, there are significant disparities in both the spectral range and DOS between the two geometries. 

We then show the spectral DOS extracted from the potential formulation [See Eq. (\ref{2d_pot})] in Figs. \ref{fig5}(c1)(c2) for the two shapes. The spectral structure is in perfect agreement with the energy spectra from exact diagonalization in Figs. \ref{fig5}(b1)(b2), demonstrating a clear geometric dependence. For the rhombus geometry, the procedure begins with a basis transformation as illustrated in Fig. \ref{fig3}, followed by the analytical continuation of the new momenta, i.e., $H(k_x,k_y) \rightarrow H(k_1',k_2') \rightarrow H(\beta_1',\beta_2')$, and then calculating the spectral potential defined in Eq. (\ref{2d_pot}). The final step is solving the Poisson equation. To rigorously validate that our formula yields the non-Bloch spectra in the TDL, we examine the deviation between the spectral potential $\phi(E)$ obtained from exact diagonalization [defined in Eq. (\ref{def_pot1})] and $\Phi(E)$ predicted by our formulation in Eq. (\ref{2d_pot}). We take several representative spectral regions [see colored boxes in Fig. \ref{fig5}(c1)(c2)] and perform a finite-size analysis of the average absolute deviation $|\phi(E)-\Phi(E)|$ in each region. For each boxed region, 49 energy points are chosen. We have also analyzed the average deviation over the entire complex plane. As shown in Figs. \ref{fig5}(d1)(d2), for both the square and rhombus, the deviations tend to zero in the TDL. Thus, the numerical spectra converge to the non-Bloch spectra predicted by our formulation.

\section{Generalized Brillouin zone} \label{seciv}
The generalized Brillouin zone (GBZ) is a concept in non-Hermitian physics generalized from the Brillouin zone \cite{nhse1}. It simultaneously encodes the localization information of skin modes and the non-Bloch spectra in the TDL. This implies that the formulation of the GBZ requires the convergence and stability of the non-Bloch spectra as prerequisites. However, spectral convergence is not guaranteed when scale-free localization occurs. Therefore, the construction of the GBZ should stick to cases with convergent non-Bloch spectra.

Strictly speaking, the GBZ is the solutions of the ChP $f(\beta, E)=0$ when the reference energies are taken from the non-Bloch spectra. In 1D, the ChP is a Laurent series with two complex variables. For a given $E$ inside the non-Bloch spectra, the solutions of the ChP are isolated points. As the non-Bloch spectra form arcs in 1D, sweeping $E$ across the entire non-Bloch spectra traces closed trajectories, i.e., the GBZ in the complex-$\beta$ plane. According to the GBZ condition Eq. (\ref{gbz1d}), these trajectories are formed by the two intermediate solutions with the same modulus. The localization length of skin modes with eigenenergy $E$ is given by $\frac{1}{\log|\beta_p(E)|}$. From the 1D case, we can extend the GBZ to $d>1$D. Namely, the GBZ consists of the $(\beta_1, \beta_2, \cdots, \beta_d)$-solutions of the ChP $f(\beta_1, \beta_2, \cdots, \beta_d, E)=0$, where $\log|\beta_j|$ is the inverse localization length along the $j$-th direction of the skin mode with eigenenergy $E$.

The geometry-dependence of the non-Bloch spectra dictates that the GBZ is also geometry-dependent. The existence of $d+1$ complex variables in the ChP makes it daunting to analytically obtain the GBZ via boundary equations except in 1D. However, in our formulation, there is a natural choice of the GBZ. Note that the local spectral potential in Eq. (\ref{dd_pot}) comes from the set $\{\tilde{\Phi}_1(E),\tilde{\Phi}_2(E),\cdots,\tilde{\Phi}_d(E)\}$. Each $\tilde{\Phi}_j(E)$ involves a minimization of spectral potential for an intermediate cylindrical system with respect to $\mu_j$. This procedure determines the inverse localization length along the $j$-th direction, denoted as $\mu_{j,\text{min}}~(j=1,2,\cdots,d)$. Once these inverse localization lengths are fixed, the associated lattice momenta $(k_1,k_2,\cdots,k_d)$ are determined by the ChP $f(e^{i k_1+\mu_{1,\text{min}}}, e^{i k_2+\mu_{2,\text{min}}},\cdots, e^{i k_d+\mu_{d,\text{min}}},E)=0$, labeled as $(k_{1,\text{min}},k_{2,\text{min}},\cdots,k_{d,\text{min}})$. Thus, we can write the GBZ condition as: 
\begin{eqnarray}\label{dd_gbz}
 \{\mu_{1,\text{min}},\mu_{2,\text{min}},\cdots,\mu_{d,\text{min}};k_{1,\text{min}},k_{2,\text{min}},\cdots,k_{d,\text{min}}\}.\notag\\
\end{eqnarray}
This GBZ is associated with the regular geometry specified by the lattice momenta $(k_1,k_2,\cdots,k_d)$. For another geometry, one needs to perform basis transformations on the Hamiltonian.

\begin{figure}
	\includegraphics[width=3.375in]{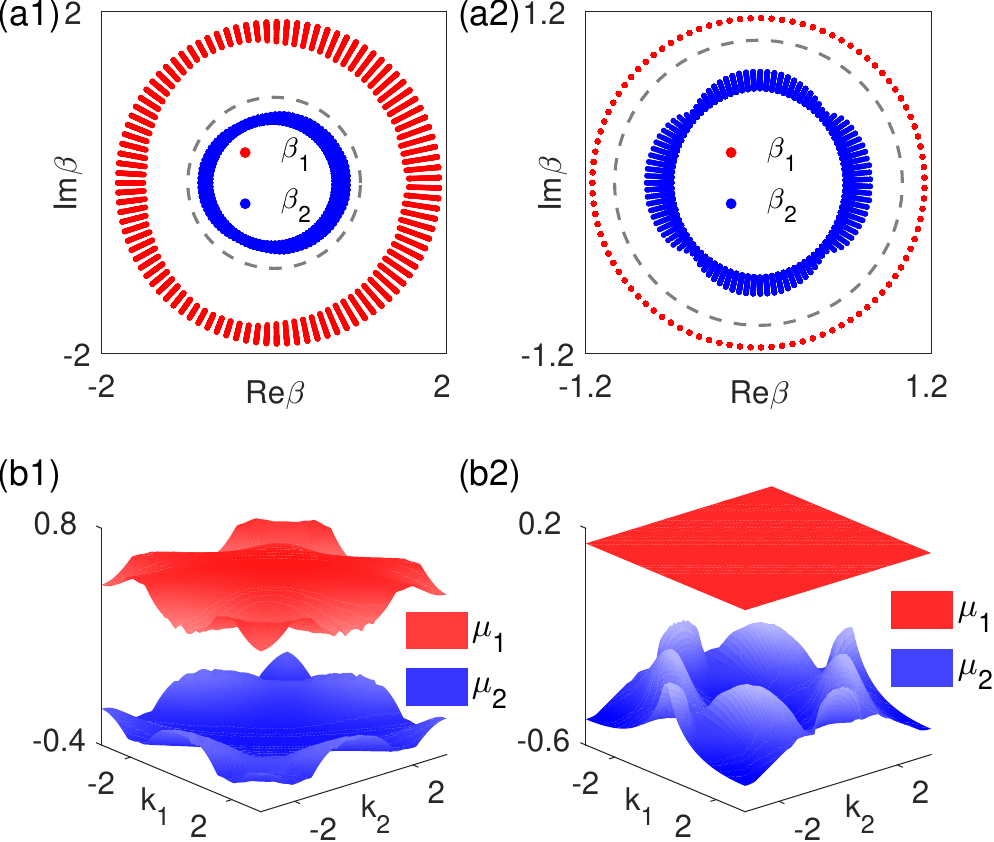}
	\caption{\label{fig6}Geometry-dependent GBZ of the Hamiltonian in Eq. (\ref{cse_model}). The GBZ for a given geometry is determined by Eq. (\ref{dd_gbz}). (a1)(a2) The GBZ visualized in the complex-$\beta$ plane for the square and rhombus shapes, respectively. The Brillouin zone (dashed gray unit circle) is included for reference. (b1)(b2) The GBZ visualized as 2D $\mu$-surfaces with explicit momentum dependence for the two geometries. Here, $\mu_j$ is the inverse localization length along the cut-$j$ direction, $\mu_j = \log|\beta_j|$ ($j=1,2$).}
\end{figure}
Let us continue with model (\ref{cse_model}) as example. The GBZ can be visualized either in the complex-$\beta$ plane or as 2D surfaces $\mu_{j,\text{min}}(k_1,k_2)$ ($j=1,2$) with explicit momentum dependence. Figure \ref{fig6} illustrates the GBZ for both square and rhombus geometries. For either shape, the GBZ in the $\beta$-plane occupies a finite region around the origin [See Fig. \ref{fig6}(a1)(a2)]. This contrasts sharply with the 1D case, where the GBZ forms 1D closed loops. In the square geometry, we have $|\beta_2|<1<|\beta_1|$, or equivalently, $\mu_2<0<\mu_1$, as shown in Fig. \ref{fig6}(b1). This indicates that skin localization occurs at the lower right corner, consistent with the spatial distributions of the skin modes in Fig. \ref{fig5}(a1). For the rhombus shape, a basis transformation to the Bloch Hamiltonian $H(k_x,k_y)$ is required before applying the GBZ condition in Eq. (\ref{dd_gbz}). As shown in Fig. \ref{fig6}(a2), we also have $|\beta_2|<1<|\beta_1|$, which is consistent with skin localization at the right corner as depicted in Fig. \ref{fig5}(a2). Note that in this case, the lattice cuts are along the $(1,1)$ and $(-1,1)$ directions. Specifically, the $\beta_1$ trajectory forms a closed circle outside the Brillouin zone, signifying a uniform localization length along the cut-1 (i.e., the (1,1) direction) for all skin modes. Consequently, the $\mu_1$-surface is flat, as shown in Fig. \ref{fig6}(b2).

\section{Spectral relations} \label{secv}
In the previous section, we demonstrated the geometry-dependence of non-Bloch spectra in the TDL, focusing exclusively on regular lattice geometries. In this section, we discuss how to treat irregular lattice shapes and explore the Amoeba formulation \cite{wz} of non-Bloch bands. In the Amoeba formulation, the spectral potential is conjectured through a minimization of Ronkin's function and the geometric information is not involved. Following a pedagogical introduction to the Amoeba formulation, we will elucidate its physical implications and highlight that Amoeba spectra generally \textit{do not} align with non-Bloch spectra of regular lattice geometries in the TDL. Moreover, we will clarify the relationships between various types of OBC energy spectra and propose a conjecture that Amoeba spectra represent the union of non-Bloch spectra for all possible lattice geometries.

\subsection{Amoeba formulation} \label{secva}
In our formulation, the geometric information is input through basis transformation, and the spectral potential under a specific geometry is derived hierarchically, as in Eqs. (\ref{2d_pot}) and (\ref{dd_pot}). The Amoeba formulation, on the other hand, starts by extending the GBZ condition Eq. (\ref{gbz1d}). For a given $E$, the ChP $f(E,\beta_1,\cdots,\beta_d)=0$ has infinite roots, forming an amoeba in the $d$D space spanned by $(\log|\beta_1|,\log|\beta_2|,\ldots,\log|\beta_d|)$. In higher dimensions, the degenerate $\beta$-solutions in the condition are generalized to be the absence of a \textit{central hole} in the amoeba. The central hole is a region in the amoeba where the winding numbers along all $d$ directions vanish. That is, if $(\alpha_1,\cdots,\alpha_d) \in \textrm{central hole}$, then
\begin{eqnarray}\label{1dwinding_A}
w_j(E)=&& \int^{2\pi}_0 \frac{d k_j}{2\pi i} \partial_{k_j} \log f(e^{i k_1+\alpha_1},\cdots,e^{i k_d+\alpha_d},E)=0,\notag\\
\end{eqnarray}
for $j=1,2,...,d$. If, for a given $E$, the associated amoeba roots lack a central hole, then $E$ is included in the Amoeba spectra, denoted as $\sigma_{\text{Amoeba}}$. Formally,
\begin{eqnarray}\label{amoeba_spectra}
 \sigma_{\text{Amoeba}} := \{\textrm{Any}~E~\textrm{without a central hole}\}. 
\end{eqnarray}
 It is important to note that the above definition only determines the spectral range. To extract the DOS, one must know the spectral potential and solve the Poisson equation (\ref{po_eq}). Wang et al. conjectured the following construction of the spectral potential \cite{wz}: 
\begin{eqnarray}\label{amoebap}
\Phi_{\text{Amoeba}}(E) = \min_{\bm{\mu}}\int \frac{d^d\bm{k}}{(2\pi)^d} \log|f(e^{i\bm{k}+\bm{\mu}},E)|, 
\end{eqnarray}
where $\bm{k} = (k_1, \cdots, k_d)$ and $\bm{\mu} = (\mu_1, \cdots, \mu_d)$. This can be seen as a direct generalization of the 1D potential function in Eq. (\ref{1d_potential}) to higher dimensions. 

It can be strictly proven that: (i) The Amoeba potential is geometry-irrelevant. Under basis transformations, $\Phi_{\text{Amoeba}}(E) $ remains unchanged. (ii) The Amoeba spectra $\sigma_{\text{Amoeba}}$ are equivalent to the uniform spectra $\sigma_{\text{uniform}}$ [See Appendix \ref{appendixc} for an introduction] obtained by eliminating the point-gaps in all directions \cite{huuniform}. This further highlights the geometry independence. (iii)  $\Phi_{\text{Amoeba}}(E)$ is always greater than or equal to the spectral potential in Eqs. (\ref{2d_pot})(\ref{dd_pot}) of our formulation. To visualize this, let us consider the 2D case:
\begin{eqnarray}\label{phiandPhi}
&&\Phi_{\text{Amoeba}}(E)\notag\\
&&= \iint\frac{d k_1 d k_2}{(2\pi)^2}\log|f(e^{i k_1+\mu_{1,\text{min}}}, e^{i k_2+\mu_{2,\text{min}}},E)|\notag\\
&&\geq \int\frac{d k_1}{2\pi}\min_{\mu_2}\left[\int\frac{d k_2}{2\pi}\log|f(e^{i k_1+\mu_{1,\text{min}}}, e^{i k_2+\mu_2},E)|\right]\notag\\
&&\geq \min_{\mu_1}\int\frac{d k_1}{2\pi}\Phi_{k_1-i\mu_1}= \bar{\Phi}_1(E)\notag\\
&&\geq \Phi(E).
\end{eqnarray}
The second line is the definition, with $\mu_{1,\text{min}}, \mu_{2,\text{min}}$ the values minimizing the Amoeba potential. In the fourth line, we have substituted the second integral with Eq. (\ref{1d_potential}) and the last line is due to Eq. (\ref{2d_pot}).

\subsection{Amoeba spectra vs non-Bloch spectra} \label{secvb}
\begin{figure}[!t]
\centering
\includegraphics[width=3.375 in]{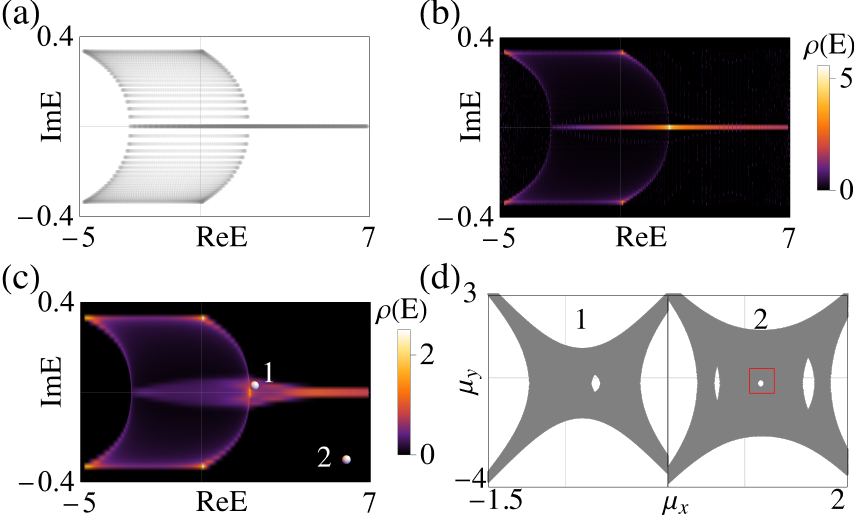}
\caption{Comparison between the Amoeba spectra and the non-Bloch spectra for model (\ref{fc_model}) under square geometry. (a) Eigenspectra from exact diagonalization with lattice size $N=5625$. (b) The spectral DOS obtained from our potential formalism in Eq. (\ref{2d_pot}) or exact solution in Appendix \ref{appendixb}. (c) The spectral DOS obtained from the Amoeba potential in Eq. (\ref{amoebap}). (d) The amoeba solutions of the ChP for two reference energies $E_1=2.2+0.03i$, $E_2=6-0.3i$, marked by white dots in (c). The central hole (inside the red box) only exists for $E_2$. The hole for $E_1$ in the left panel has a nonzero winding number, thus not a central hole.}\label{fig7}
\end{figure}

We compare the Amoeba formulation and our potential formulation through an analytically tractable case from Ref. \cite{fangchen}. The Hamiltonian reads:
\begin{eqnarray}\label{fc_model}
H(k_x,k_y)=&&[5(\cos k_x + \cos 2k_x) - i(\sin k_x + 3\sin 2k_x) \notag\\
&&+ 5\cos k_y + i\sin k_y]/2.
\end{eqnarray}
The energy spectra obtained from numerical diagonalization are shown in Fig. \ref{fig7}(a). The eigenstates are well-localized skin modes at one of the corners \cite{fangchen}. In Fig. \ref{fig7}(b), we illustrate the spectral DOS obtained from our potential formalism, which is in perfect agreement with the numerical results. Note that this model decouples in the $x$ and $y$ directions, i.e., $H(k_x,k_y)=H_x(k_x)+H_y(k_y)$, where $H_x(k_x)$ and $H_y(k_y)$ depend solely on $k_x$ and $k_y$, respectively. The non-Bloch spectra under square geometry in the TDL can be analytically derived using the 1D non-Bloch theory applied independently in each direction [See Appendix \ref{appendixb}]. In this way, the same spectral DOS can be precisely benchmarked, thereby confirming the accuracy of our potential formulation. 

Next, we present the spectral DOS extracted from the Amoeba formulation [See Eq. (\ref{amoebap})] in Fig. \ref{fig7}(c). A comparison with the exact DOS in Fig. \ref{fig7}(b) reveals noticeable differences in both the spectral range and density near the real energy axis. We examine two representative eigenenergies $E_1=2.2+0.03i$, $E_2=6-0.3i$ (marked as white dots in Fig. \ref{fig7}(c)) and showcase their amoeba-shaped solutions of the ChP in Fig. \ref{fig7}(d). For $E_2\notin\sigma_{\text{Amoeba}}$, there exists a central hole in the amoeba, whereas for $E_1\in\sigma_{\text{Amoeba}}$, no such central hole exists. This is consistent with the definition of Amoeba spectra given in Eq. (\ref{amoeba_spectra}).

\subsection{Spectral relations}\label{secvc}
Unless in 1D, the Amoeba spectra generally do not coincide with the non-Bloch spectra for non-Hermitian systems with regular lattice shapes in the TDL. Extensive model calculations indicate that (i) the energy spectra under a given shape fall within the range of $\sigma_{\text{Amoeba}}$; (ii) for systems with smooth boundaries (e.g., a disk in 2D), the spectra in the TDL tend towards $\sigma_{\text{Amoeba}}$ \cite{wz}. We first provide an exact proof of statement (i).\\
\underline{\textbf{Theorem:}}
\begin{eqnarray}
\sigma_{G_1} \neq &\sigma_{G_2} \neq \sigma_{\text{Amoeba}}, \quad \text{if} \quad G_1 \neq G_2; \label{relation1} \\
&\sigma_{G} \subseteq \sigma_{\text{Amoeba}}. \label{relation2}
\end{eqnarray}
\textit{Proof:} $\sigma_{G,G_1,G_2}$ denote the non-Bloch spectra under the geometric shape $G$, $G_1$, $G_2$, respectively. Briefly, we consider the 2D case, with straightforward generalization to higher dimensions. Eq. (\ref{relation1}) manifests itself in Fig. \ref{fig5} where the square and rhombus geometries are considered. Through the analytically tractable case in Fig. \ref{fig7}, the difference between the non-Bloch spectra for a given geometry and $\sigma_{\text{Amoeba}}$ is validated. We note that Eq. (\ref{relation1}) makes a strong assertion about the geometry-dependence of the non-Bloch spectra and their distinction from the Amoeba spectra, although cases where the non-Bloch spectra are equal to the Amoeba spectra do exist. Now let us focus on Eq. (\ref{relation2}). For any $E \notin \sigma_{\text{Amoeba}}$, there exists a central hole in its amoeba. Let us take an arbitrary point $(\alpha_1,\alpha_2)$ inside this central hole. It follows that all the equalities in Eq. (\ref{phiandPhi}) hold due to: (1) $\Phi_{\text{Amoeba}}(E)$ being independent of the choice of $(\alpha_1,\alpha_2)$ within the central hole \cite{wz}, and (2) the vanishing of spectral winding along any direction for the central hole [See Eq. (\ref{1dwinding_A})]. Thus, $\alpha_1,\alpha_2$ minimize the integrals. That is, for any $E \notin \sigma_{\text{Amoeba}}$, $\Phi(E)=\Phi_{\text{Amoeba}}(E)$. The spectral DOS is governed by the $\rho(E)=\frac{1}{2\pi}\nabla^2\Phi(E)$. Since $\nabla^2\Phi_{\text{Amoeba}}(E)=0$ for $E \notin \sigma_{\text{Amoeba}}$, it follows that $E \notin \sigma_G$. Hence, $\sigma_{G} \subseteq \sigma_{\text{Amoeba}}$. \qed

\begin{figure}[!t]
\centering
\includegraphics[width=2.1 in]{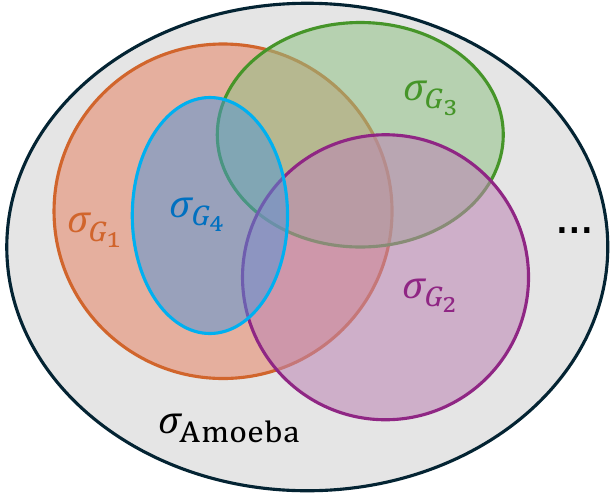}
\caption{Sketch of spectral relations in higher dimensions. $\sigma_{\text{Amoeba}}$: Amoeba spectra defined in Eq. (\ref{amoeba_spectra}) through the absence of amoeba central hole. $\sigma_{G_1},\sigma_{G_2},...$ represent the non-Bloch spectra associated with lattice geometry $G_1,G_2,...$ in the TDL.}\label{fig8}
\end{figure}
Statement (ii) is difficult to prove rigorously. We examine it from a physical perspective and summarize it as\\
\underline{\textbf{Conjecture-1:}}
\begin{eqnarray}
\bigcup_{G} \sigma_{G} &=& \sigma_{\text{Amoeba}}; \label{relation4}\\
\sigma_{\text{Smooth}} &=& \sigma_{\text{Amoeba}}. \label{relation3}
\end{eqnarray}
Here, $\sigma_{\text{Smooth}}$ denotes the non-Bloch spectra associated with smooth lattice shapes, such as a disk or ball in 2D or 3D. This conjecture should be understood in a set-theoretical sense and the spectral relations are sketched in Fig. \ref{fig8}. While different geometries have distinct non-Bloch spectra, these spectra all lie within $\sigma_{\text{Amoeba}}$, which represents the union of the non-Bloch spectra across all possible geometries.

A handwaving argument for the relations above proceeds as follows. In the Amoeba formulation, geometry information is not a required input. However, for systems with regular lattice cuts, we must employ the spectral potential in Eq. (\ref{dd_pot}) to derive their non-Bloch spectra in the TDL. This spectral potential depends on geometry but is never greater than $\Phi_{\text{Amoeba}}(E)$, as stipulated by Eq. (\ref{phiandPhi}). Under regular lattice geometries, the system automatically minimizes its spectral potential in a manner better than Amoeba. For $E \notin \sigma_{\text{Amoeba}}$, $\Phi(E)=\Phi_{\text{Amoeba}}(E)$ holds for any regular shape $G$. Specifying a lattice geometry only reduces the spectral potential within $\sigma_{\text{Amoeba}}$ and alters the DOS accordingly. For a given shape $G$, the deviation from the Amoeba potential occurs in specific regions (referred to as shift regions) within $\sigma_{\text{Amoeba}}$, impacting the DOS solely in those areas. Different lattice shapes possess distinct shift regions. By sweeping over all possible lattice geometries, the unshifted regions would cover the entirety of $\sigma_{\text{Amoeba}}$. Therefore, the union of all possible non-Bloch spectra $\sigma_G$ should coincide with $\sigma_{\text{Amoeba}}$ in a set-theoretic sense. In other words, $\sigma_{\text{Amoeba}}$ encompasses contributions from all possible geometries. For smooth shapes, where lattice cuts can be oriented in any direction, the non-Bloch spectra in the TDL should tend toward $\sigma_{\text{Amoeba}}$, as suggested by numerical studies \cite{wz}.

The discussion above suggests a promising strategy for handling complex geometric shapes like irregular polygons in 2D \cite{huuniform}. For stable non-Bloch spectra, the eigenenergies and eigenstates in large systems should closely match the non-Bloch spectra and skin modes in the TDL. The appearance of skin modes is related to how the edges of the polygons are connected. Therefore, we can break down the irregular shapes into regular parallelograms or parallelepipeds in 2D or 3D, respectively. The non-Bloch spectra in the TDL come from all subsystems, to which our non-Bloch band theory developed in Section \ref{seciii} applies. We expect the spectral potential to be a combination of the spectral potentials for each regular shape, following the guiding minimization principle. We will address these complex cases in future works.

\section{Critical NHSE} \label{secvi}
The formulation of the non-Bloch band theory assumes that the system's eigenvalues and skin modes stabilize when the system size is sufficiently large. However, this assumption does not always hold when the skin modes exhibit scale-free localization. In this section, we consider such a scenario, dubbed higher-dimensional critical NHSE, where the skin modes reside on the boundaries and strongly depend on system size. Notably, a special type of skin effect exists in 1D, namely the 1D critical NHSE \cite{llhcritical}, where the system's spectra and localization behavior of skin modes show high sensitivity to system size. We will reveal strong similarities between the higher-dimensional and 1D critical NHSE, which justifies the term ``critical". Drawing insights from the 1D case, we will identify the conditions under which the critical NHSE occurs. We will demonstrate that for geometries with regular lattice cuts, the non-Bloch spectra are not well-defined in the TDL. Additional geometric details, such as boundary ratios, significantly affect the system's eigenspectra. A detailed discussion on the spectral instability in the presence of weak perturbation will be left to Section \ref{secviii}.

\subsection{Insights from the 1D case} \label{secvia}
\begin{figure}[!t]
\centering
\includegraphics[width=3.375 in]{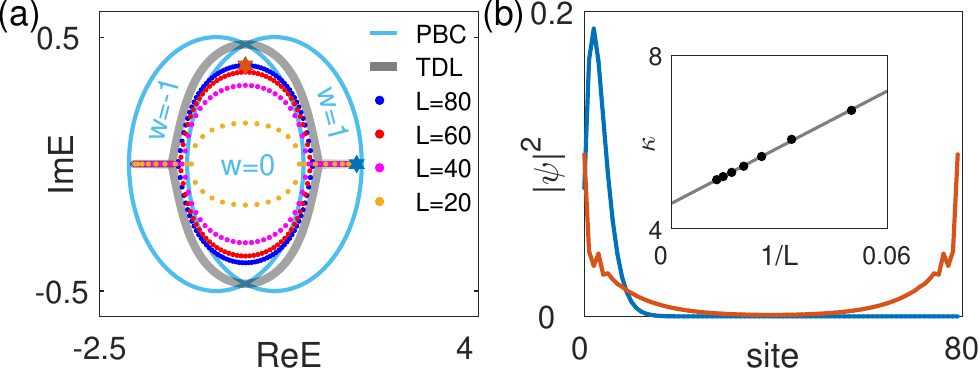}
\caption{Illustration of the 1D critical NHSE. (a) Energy spectra of the model (\ref{1dcnhse_model}). The PBC spectra are shown in cyan. The OBC spectra for different lattice lengths $L=20,40,60,80$ (of a single chain) are shown in orange, magenta, red, and blue, respectively, with their non-Bloch spectra in the TDL shown in grey. (b) Spatial distributions of the two eigenstates (marked by colored stars in (a)) with the largest real and imaginary parts. (Inset) The dependence of the localization length of the chosen eigenstate in the central region with respect to system length. The fitting line is shown in grey. $t_{1L}=t_{2R}=0.5$, $t_{1R}=t_{2L}=1$, $V=0.5$, $\delta=0.01$.}\label{fig9}
\end{figure}

Let us revisit the 1D critical NHSE using the following prototypical model \cite{llhcritical}:
\begin{eqnarray}\label{1dcnhse_model}
H= \begin{pmatrix}
t_{1L}\beta^{-1} + t_{1R}\beta+V & \delta \\
\delta & t_{2L}\beta^{-1} + t_{2R}\beta-V
\end{pmatrix}.
\end{eqnarray}
It is a double-chain Hatano-Nelson model \cite{hnmodel} with inter-chain coupling of strength $\delta$. $t_{1L}$ or $t_{1R}$ ($t_{2L}$ or $t_{2R}$) is the hopping to the left or right for the first (second) chain. We set $t_{1L}=t_{2R}=0.5$, $t_{1R}=t_{2L}=1$, $V=0.5$. In the absence of inter-chain couplings, the two chains exhibit opposite skin localizations. We take a very weak coupling strength $\delta = 0.01$. The energy spectra for different system sizes $L = 20, 40, 60, 80$ are depicted in Fig. \ref{fig9}(a). We can observe that the central part of the spectra is size-dependent and tends towards the non-Bloch spectra (grey curves) predicted by the potential formalism in Eq. (\ref{1dlocal}) in the TDL. However, the two sides of the spectra stay intact as the system size $L$ increases. Next, we examine two representative eigenstates, whose eigenenergies have the largest real or imaginary parts, respectively. Fig. \ref{fig9}(b) plots the spatial distributions of the two eigenstates for a system size $L=80$. The first eigenstate is well-localized at the left boundary, while the second exhibits a size-dependent localization. This is visualized in the inset, which plots the localization lengths of the eigenstates with respect to varying system sizes. The localization lengths are extracted by fitting the spatial profiles of the eigenstates to an exponential function. This is consistent with their spectral instability as the system size increases.

To understand the distinct localization behaviors of the two eigenstates and the spectral instability as the system size increases, we consider the spectral winding number of the Bloch Hamiltonian. For a reference energy $E$, the winding number is defined as:
\begin{eqnarray}\label{1dwinding} 
w(E) = \int^{2\pi}_0 \frac{dk}{2\pi i} \partial_{k} \log\det[H(k)-E]. \end{eqnarray}
In 1D, $w(E)$ is an integer topological invariant that counts the spectral windings of the loop-shaped Bloch spectra with respect to the reference energy $E$. A non-zero spectral winding implies the presence of point gap, which is the topological origin of NHSE \cite{nhse6,huuniform}. The correspondence between the spectral winding number and skin modes has been well established in 1D \cite{nhse5}.

For the model (\ref{1dcnhse_model}), let us examine the winding number of its Bloch Hamiltonian $H(k)$. We have $w(E)=\pm 1$ if $E$ lies at the two wings [See Fig. \ref{fig9}(a)]. The ``winding number-skin mode" correspondence then dictates that the eigenstates at the two wings should be normal skin modes exponentially localized at the boundaries. When $E$ falls within the central region, we have $w(E)=0$. This aligns with the size-dependent localization of eigenstates there. Note that the model (\ref{1dcnhse_model}) is a two-band system, with its Bloch spectra comprising two loops. The winding number sums the net contributions of the two bands and can be expressed as $w(E)=w^{\text{band-1}}(E)+w^{\text{band-2}}(E)$, where $w^{\text{band-j}}(E)=\int^{2\pi}_0\frac{dk}{2\pi i} \partial_{k} \log(E_j-E)$ $(j=1,2)$. In the central region, $w^{\text{band-1}}=-w^{\text{band-2}}=-1$; whereas at the left or right wing, $w^{\text{band-1}}=-1, w^{\text{band-2}}=0$ or $w^{\text{band-1}}=0, w^{\text{band-2}}=1$. Thus, the size-dependent spectra and skin modes are tied to the zero net winding in the central region. Now, we are ready to explore the higher-dimensional extensions of the critical NHSE.

\subsection{Net winding numbers} \label{secvib}
For a general $d$D Bloch Hamiltonian $H(k_1,\cdots,k_d)$, its Brillouin zone forms a $d$D torus. The winding number along the $j$-th direction with respect to a reference energy $E$, is defined as:
\begin{eqnarray}
w_j(E)=\int^{2\pi}_0\frac{dk_j}{2\pi i} \partial_{k_j} \log\det[H(k_1,\cdots,k_d)-E],
\end{eqnarray}
with $j=1,\cdots,d$. It is a function of all the momenta except $k_j$. The net winding number is the sum of $w_j$ over all the other momenta, namely,
\begin{eqnarray}
\bar{w}_j(E)=\int^{2\pi}_0\frac{dk_1\cdots dk_d}{i(2\pi)^d} \partial_{k_j} \log\det[H(k_1,\cdots,k_d)-E].\notag\\
\end{eqnarray}
For instance, in 2D, the net winding number along the $k_x$ direction is given by $\bar{w}_{x}=\int^{2\pi}_0\frac{dk_x dk_y}{i(2\pi)^2} \partial_{k_x} \log\det[H(k_x,k_y)-E]$. Similarly, the net winding number along the $k_{\theta}$ direction is expressed as $\bar{w}_{k_{\theta}}=\int^{2\pi}_0\frac{dk_{\theta}dk_{\perp}}{i(2\pi)^2} \partial_{k_{\theta}} \log\det[H(k_{\theta},k_{\perp})-E]$, where $k_{\perp}$ represents the momentum normal to $k_{\theta}$. 

The net winding number quantifies the average spectral winding along a specific direction in the Brillouin zone and may not take an integer value. Notably in $d$D, if there exist $d$ independent directions with vanishing net winding numbers, then along any other direction, the net winding number should be zero. Thus, the net winding number allows for a complete and exclusive classification of NHSE in higher dimensions, which will be detailed in Section \ref{secix}. For model (\ref{cse_model}), when the reference energy is chosen inside the non-Bloch spectra, the net winding numbers along the $x$, $y$, $(1,1)$, and $(-1,1)$ directions are nonzero, resulting in corner skin modes. This can be regarded as a higher-dimensional generalization of the ``winding number-skin mode" correspondence in 1D. And the natural extension of the 1D critical NHSE to higher dimensions should be the vanishing net winding numbers along all directions. In the following two subsections, we will unveil the universal features of higher-dimensional critical NHSE through a concrete model.

\begin{figure}[!t]
\centering
\includegraphics[width=3.375 in]{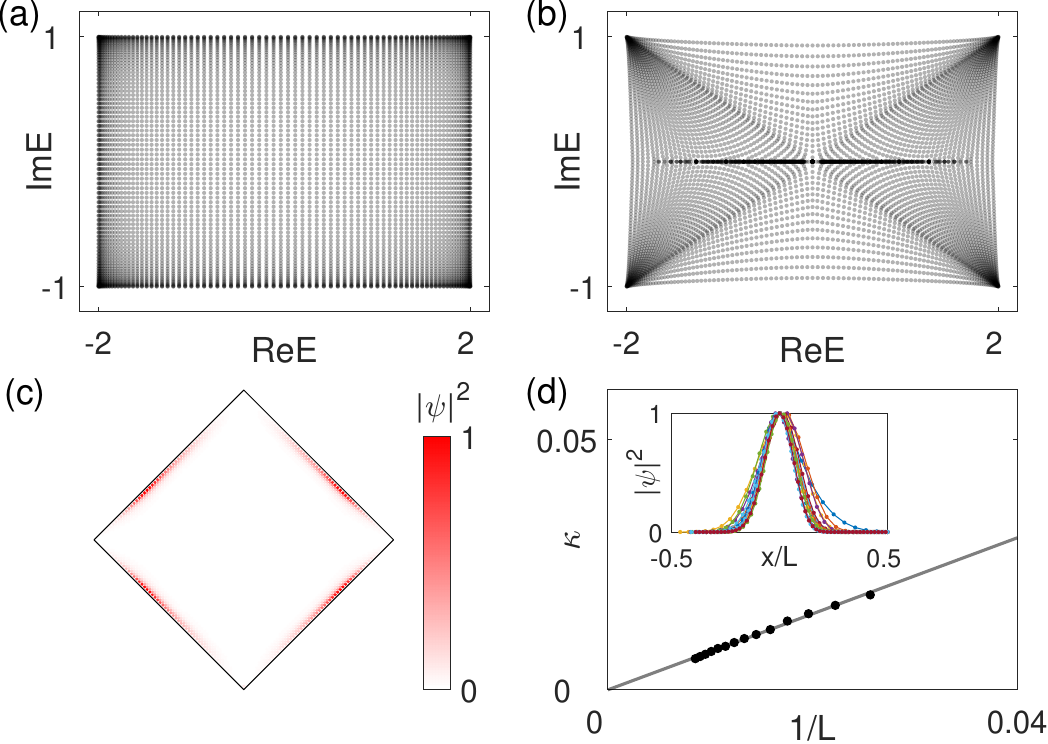}
\caption{Scale-free localization in 2D critical NHSE. (a)(b) Eigenspectra from exact diagonalization of model (\ref{gdse_model}) on the square (with system size $N=6400$) and rhombus geometry (with system size $N=6385$), respectively. (c) Spatial distributions of all eigenstates under rhombus geometry. (d) Finite-size analysis of the eigenstates’ broadening extracted from Gauss fitting (inset). The inverse localization length $\kappa$ scales linearly with $1/L$ and extrapolates to zero (grey line) in the TDL. $L$ is the boundary length.}\label{fig10}
\end{figure}
\subsection{Scale-free localization} \label{secvic}
We take the 2D model from Ref. \cite{fangchen} to illustrate the critical NHSE. The Hamiltonian is
\begin{eqnarray} \label{gdse_model}
H(k_x,k_y)=2\cos k_x+i\cos k_y.
\end{eqnarray}
The $x$ and $y$ directions are decoupled for this model. For a fixed value of either $k_x$ or $k_y$, the Bloch spectrum forms a straight line along either the imaginary or real energy axis. It is evident that $\bar{w}_x(E)=\bar{w}_y(E)=0$ for any reference $E$. Thus, the model meets the criterion for critical NHSE. The OBC energy spectra for square and rhombus geometries are shown in Figs. \ref{fig10}(a)(b). Although their spectral ranges appear nearly identical, their spectral DOS shows a clear distinction. For the square/rhombus, more eigenenergies are observed at the boundary region/diagonal lines and the real energy axis of the spectra. Figure \ref{fig10}(c) plots the spatial profiles of all eigenstates under rhombus geometry. They are localized at the boundaries with finite broadening. This is different from the square geometry, where all eigenstates are extended plane waves \cite{fangchen} due to the separability of the $x$ and $y$ directions.

To highlight the key difference between these boundary-localized eigenstates and corner skin modes or extended plane waves, we vary system sizes and focus on one of the rhombus edges. We examine the relationship between the wave-packet broadening along the boundary and the boundary length $L$, as shown in Fig. \ref{fig10}(d). Here, the wave-packet width $\frac{1}{\kappa}$ was extracted by fitting the eigenstates from numerical diagonalization with a Gaussian function. We find that the width depends linearly on $L$ and tends to infinity in the TDL. Therefore, we dub these eigenstates critical or scale-free skin modes. They are in sharp contrast to either the corner skin modes or extended plane waves, whose broadenings are irrelevant to system size. For other models (e.g., the model in Ref. \cite{yzsedge}), we also find that as long as the net winding numbers are zero in all directions, the eigenstates exhibit scale-free localization at the boundaries.

\subsection{Indefiniteness of non-Bloch spectra}  \label{secvid}
A physical consequence of skin modes' scale-free localization is the indefiniteness of non-Bloch spectra in the TDL. Note that in the formulation of non-Bloch band theory, geometric information is specified by lattice-cut orientations without involving their boundary ratios. In the normal case, skin modes are well-localized with definite localization length. The energy spectra should stabilize for large system sizes and converge to the non-Bloch spectra in the TDL, regardless of these ratios. However, when scale-free localization occurs, eigenstates broaden as the boundary length increases. The energy spectra do not stabilize for large system sizes and are sensitive to the boundary ratios.
\begin{figure}[!t]
\centering
\includegraphics[width=3.375 in]{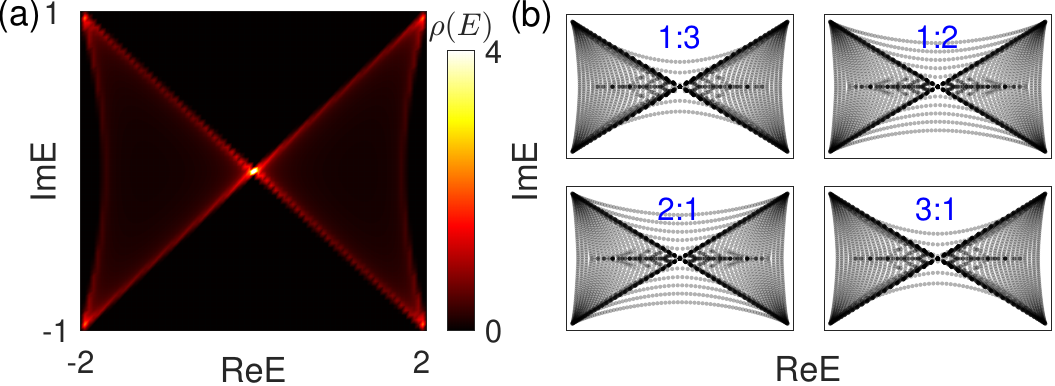}
\caption{Boundary-ratio dependence of energy spectra for the critical NHSE. (a) Spectral DOS of model (\ref{gdse_model}) under rhombus geometry extracted from our potential formalism. (b) Numerical energy spectra for four boundary ratios $3:1$, $2:1$, $1:2$, and $1:3$. The boundaries are along the $(1,1)$ and $(-1,1)$ directions. The total lattice sites are $N=3267,3278,3278,3267$, respectively.}\label{fig11}
\end{figure}

For the model (\ref{gdse_model}), we examine the boundary-ratio dependence of the energy spectra under rhombus geometry, where scale-free localization occurs. In Fig. \ref{fig11}, we showcase the DOS extracted from our potential formulation in Fig. \ref{fig11}(a). Notable discrepancies exist in the spectral range compared to the numerical data in Fig. \ref{fig10}(b). In Figs. \ref{fig11}(b)-(d), we plot the energy spectra for four different ratios, $3:1$, $2:1$, $1:2$, and $1:3$, respectively. It's evident that the energy spectra are sensitive to the ratios, despite the fixed lattice-cut along $(1,1)/(-1,1)$ directions and nearly identical lattice-site numbers. These numerical results indicate that for non-Hermitian systems hosting critical NHSE, the OBC energy spectra are not well-defined in the TDL by solely fixing the lattice-cut directions. In fact, the energy spectra for such critical systems are highly unstable when random disorder is added. A detailed discussion on the spectral instability in the presence of weak perturbations will be left to Section \ref{secviii}.

\section{Non-reciprocal NHSE}  \label{secvii}
The critical NHSE corresponds to vanishing net winding numbers in all directions. The critical case is relatively simple, with all eigenstates exhibiting scale-free localization. In contrast, there are cases where the net winding numbers are non-zero in certain directions. Due to the non-reciprocity along these directions, they are referred to as non-reciprocal NHSE. The skin modes in non-reciprocal NHSE can exhibit various types of localization, making it more complex. In the previous examples (Eq. (\ref{cse_model}) and Eq. (\ref{fc_model})), we have already seen the normal case with well-localized corner skin modes. Additionally, non-reciprocal systems may host anomalous corner modes, edge skin modes, and scale-free skin modes. We demonstrate these cases through concrete models.

\subsection{Anomalous corner skin modes}\label{secviia} 
\begin{figure}[!t]
\centering
\includegraphics[width=3.375 in]{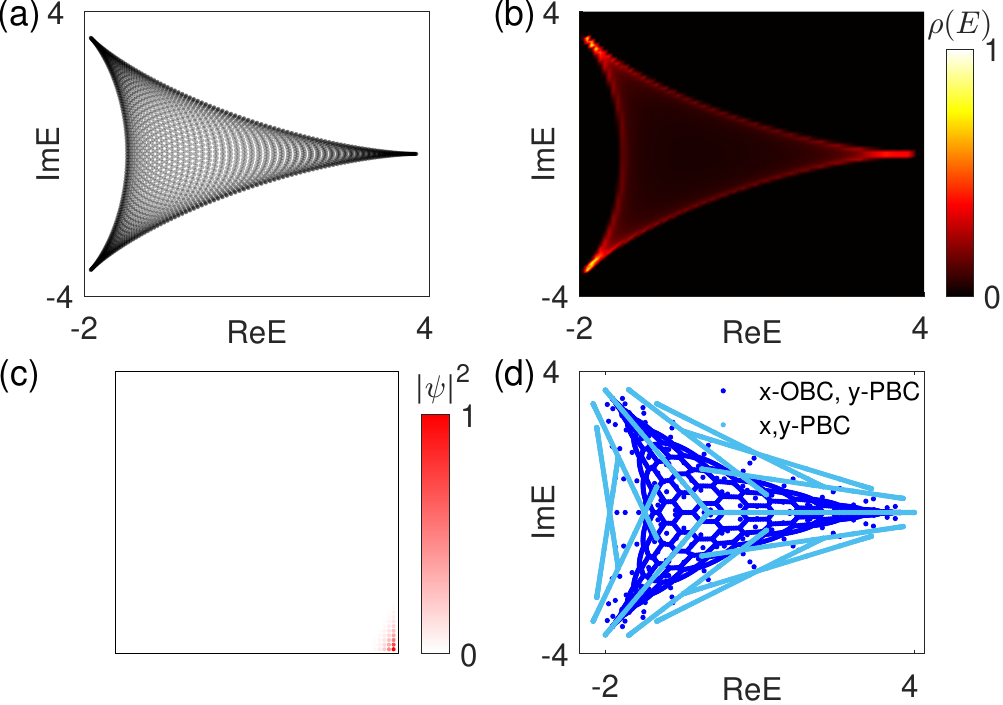}
\caption{Anomalous corner skin modes in non-reciprocal systems. (a) Energy spectra from numerical diagonalization for model (\ref{jh_model}) under square geometry with system size $N=3600$. (b) Spectral DOS obtained from our formulation in Eq. (\ref{2d_pot}), in perfect agreement with (a). (c) Spatial distributions of all eigenstates on the square geometry. (d) Energy spectra for full PBC along both $x$ and $y$ directions (in cyan) and $x$-OBC/$y$-PBC (in blue), with $L_x=15$ and $L_y=150$.}\label{fig12}
\end{figure}

The model is sourced from Ref. \cite{jianghui}, with Hamiltonian
\begin{eqnarray}\label{jh_model}
H(\beta_x,\beta_y)=2\beta_x+\beta_y+\beta_x^{-1}\beta_y^{-1}.
\end{eqnarray}
In Fig. \ref{fig12}(a), we present the energy spectra from exact diagonalization. Figure \ref{fig12}(b) shows the spectral DOS derived from our potential formulation. The perfect agreement between the numerical results and theory indicates the convergence and stability of the non-Bloch spectra in the TDL. The spatial distributions of all eigenstates are illustrated in Fig. \ref{fig12}(c), showing well-localized corner modes.

The above corner localization of the eigenstates is counterintuitive because of the absence of non-reciprocity along the $y$ direction. It's straightforward to check that $\bar{w}_y(E)=0$ for any $E$, whereas for the $x$ direction, the net winding number is non-zero. In fact, for its Bloch Hamiltonian $H(k_x,k_y)$ with fixed $k_x$, varying $k_y$ from $0$ to $2\pi$ traces line-shaped spectra in the complex plane, as shown in Fig. \ref{fig12}(d). We thus call these eigenstates anomalous corner skin modes. To understand their appearance, we note that the reciprocity condition $\bar{w}_y(E)=0$ along the $y$ direction is only satisfied when the PBC is applied along the $x$ direction. Once the OBC is applied along the $x$ direction, non-reciprocity emerges along the $y$ direction. For instance, if we consider a cylindrical geometry with OBC along the $x$ direction and PBC along the $y$ direction, spectral loops would emerge in the complex plane as $k_y$ varies from $0$ to $2\pi$, as depicted in Fig. \ref{fig12}(d). The anomalous corner accumulation thus arises from the skin localization in this cylindrical geometry.

\subsection{Edge and scale-free skin modes}  \label{secviib}
\begin{figure}[!t]
\centering
\includegraphics[width=3.375in]{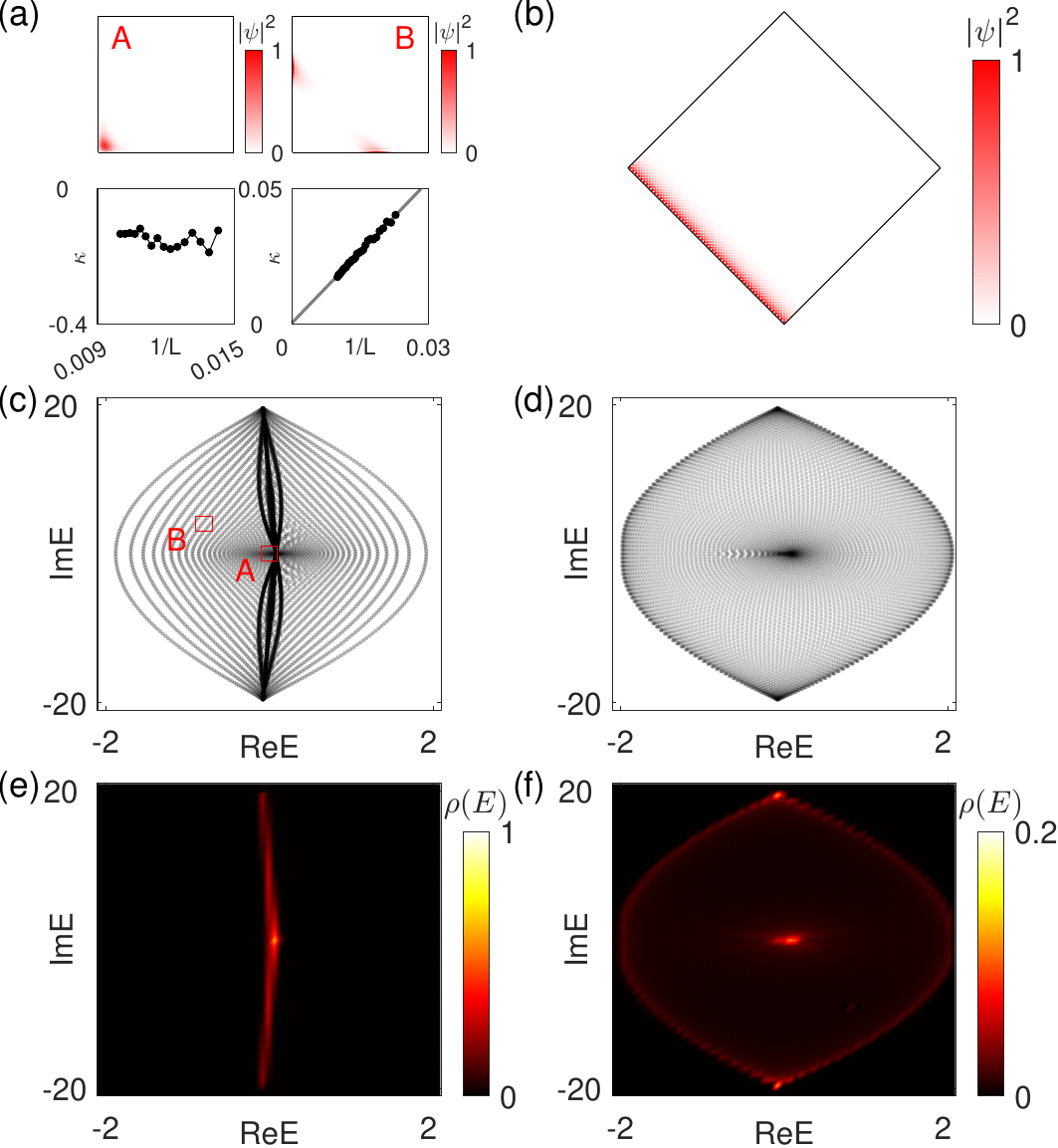}
\caption{Edge and scale-free skin modes in non-reciprocal systems. (a)(b) Spatial distributions of eigenstates on the square and rhombus geometries with system sizes $N=6400$ and $N=6385$, respectively. In (a), the eigenstates are selected from the two spectral regions A and B marked with red boxes in (c). The lower panel is finite-size analysis of eigenstates’ broadenings, with $L$ the boundary length. (c)(d) Energy spectra obtained from exact diagonalization for the two shapes. (e)(f) The spectral DOS derived from our potential formulation in Eq. (\ref{2d_pot}), demonstrating perfect agreement solely for the rhombus geometry. For the square shape, the non-Bloch spectra in the TDL do not converge.}\label{fig13}
\end{figure}

An illustrative model is given by
\begin{eqnarray}\label{am_model}
&&H(\beta_x,\beta_y)=6\beta_x-4\beta_x^{-1}+6\beta_y-4\beta_y^{-1}\notag\\
&&+\frac{1}{2}(\beta_x\beta_y+\beta_x\beta_y^{-1}+\beta_x^{-1}\beta_y+\beta_x^{-1}\beta_y^{-1}).
\end{eqnarray}
The $x$ and $y$ directions are coupled in this model. It's easy to check $\bar{w}_{(-1,1)}(E)=0$ for any $E$ due to the invariance of exchanging $\beta_x$ and $\beta_y$ in the Hamiltonian. For other directions, there exist non-reciprocity with non-vanishing net winding numbers. We examine both the square and rhombus geometries below. The manifestations of skin modes depend on the choice of the lattice shape. Figures \ref{fig13}(a)(b) depict the spatial distributions of eigenstates under square and rhombus geometries, respectively. In Fig. \ref{fig13}(c)(d), we show their corresponding energy spectra from exact diagonalization. They exhibit markedly different spectral structures. For the rhombus shape, the reciprocity condition leads to edge modes extending along the cut-$(-1,1)$ direction [See Fig. \ref{fig13}(b)]. While for the square shape, there exist two types of eigenstates, as depicted in Fig. \ref{fig13}(a), corresponding to different spectral regions as marked in Fig. \ref{fig13}(c). The eigenstates within the central region ($A$) are well-localized corner skin modes, with size-irrelevant localization lengths. The eigenstates outside the central region ($B$) are scale-free skin modes. A finite-size analysis of their broadening indicates that their localization lengths scale linearly with the boundary length. 

The example above implies that scale-free skin modes can exist in non-reciprocal systems, not just in the critical cases. Moreover, different types of skin modes can coexist within the same system. For the square shape, as boundary lengths increase, the scale-free modes extend further, which is surprising given the non-reciprocity in both the $x$ and $y$ directions. These scale-free skin modes fail the convergence of non-Bloch spectra in the TDL, and give rise to sepctral instability when changing system size or boundary ratios or introducing perturbations. In Figs. \ref{fig13}(e) and (f), we display the spectral DOS extracted from our formulation, which shows a perfect match with numerical data only for the rhombus shape. For the square shape, our theory still captures the stable part of the spectra (the central line), corresponding to the normal skin modes.

\section{Spectral instability}  \label{secviii}
This section explores the spectral instability in non-Hermitian systems. The instability caused by large exceptional degeneracies (such as large Jordan matrix) is well-known in the literature and lies outside our focus. We will investigate the spectral instability associated with various types of skin modes. Specifically, we examine how weak perturbations, such as random disorder, affect the spectral structures of non-Bloch bands in the TDL. Our analysis reveals a spectral instability stemming from the presence of scale-free skin modes. We begin with the 1D case and subsequently extend the discussions to higher dimensions.

\begin{figure}[!t]
\centering
\includegraphics[width=3.375 in]{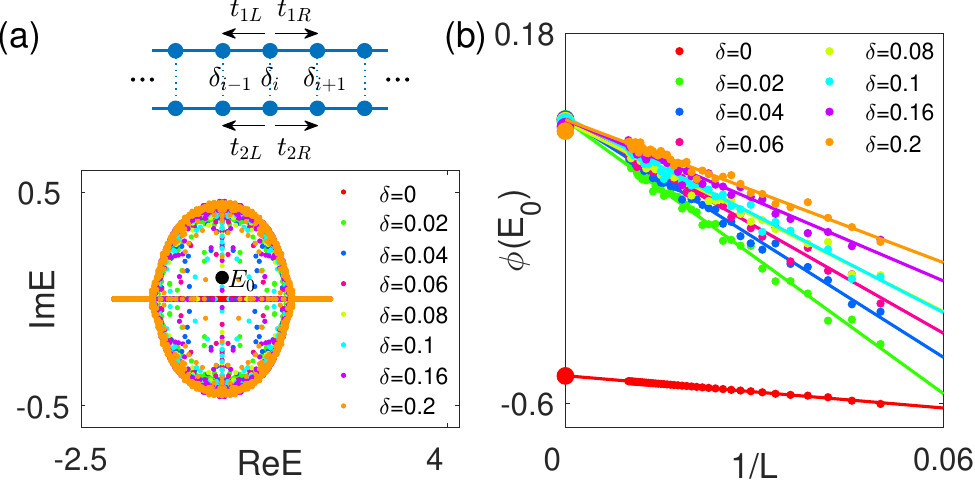}
\caption{ Effect of disorder for the 1D critical NHSE. (a) (Top) Sketch of the two-chain model. The inter-chain couplings are set to be random $\delta_i\in[-\delta,\delta]$ with $\delta$ the disorder strength. (Bottom) Energy spectra with different disorder strengths for system size $L=80$ (of a single chain). (b) Finite-size analysis of the spectral potential at $E_0=0.1i$ for different disorder strengths. $t_{1L}=t_{2R}=0.5$, $t_{1R}=t_{2L}=1$, $V=0.5$.}\label{fig14}
\end{figure}
\subsection{1D critical NHSE} \label{secviiia}
We take the 1D model (\ref{1dcnhse_model}) of critical NHSE and introduce disorder in the couplings between the two chains, as depicted in Fig. \ref{fig14}(a). The couplings are drawn from a uniform distribution of random disorder $\delta_i \in [-\delta, \delta]$ with strength $\delta$. We gradually ramp up the disorder strength and track the changes in the system's energy spectra, as shown in Fig. \ref{fig14}(a). A significant difference in the energy spectra is observed between $\delta = 0$ and $\delta \neq 0$. For the former, the two chains are fully decoupled, and the energy spectra reside on the real line. For $\delta \neq 0$, the central region of the spectra undergoes a dramatic change, corresponding to size-dependent skin modes in the clean system. In contrast, the weak disorder has a negligible effect on the spectra's two wings, demonstrating the stability of normal skin modes against disorder.

To rigorously analyze the disorder effect, we perform a finite-size analysis of the spectral potential $\phi(E_0)$ for different disorder strengths $\delta$, as shown in Fig. \ref{fig14}(b). The reference energy is chosen as $E_0 = 0.1i$. For the clean case $\delta = 0$, $\lim\limits_{L\rightarrow\infty}\phi(E_0) \approx -0.542$ in the TDL. However, for very small $\delta \neq 0$, the potential saturates to a different value $\lim\limits_{L\rightarrow\infty}\phi(E_0) \approx 0$. These two values can be determined by Eq. (\ref{1dlocal}) of the 1D potential formulation, noting that the ChP is separable for the former and non-separable for the latter. This analysis highlights the non-commutativity of the zero-perturbation limit and the TDL:
\begin{eqnarray}
\lim_{\delta\rightarrow0}\lim_{L\rightarrow\infty}\phi(E)\neq\lim_{L\rightarrow\infty}\lim_{\delta\rightarrow0}\phi(E), 
\end{eqnarray}
where the left side represents taking the TDL before the zero-perturbation limit, and the right side represents taking the TDL for the clean system. The spectral instability originating from the critical skin modes leads to the non-exchangeability of these two limits.

\subsection{Instability in higher dimensions}\label{secviiib}
\begin{figure}[!t]
\centering
\includegraphics[width=3.375 in]{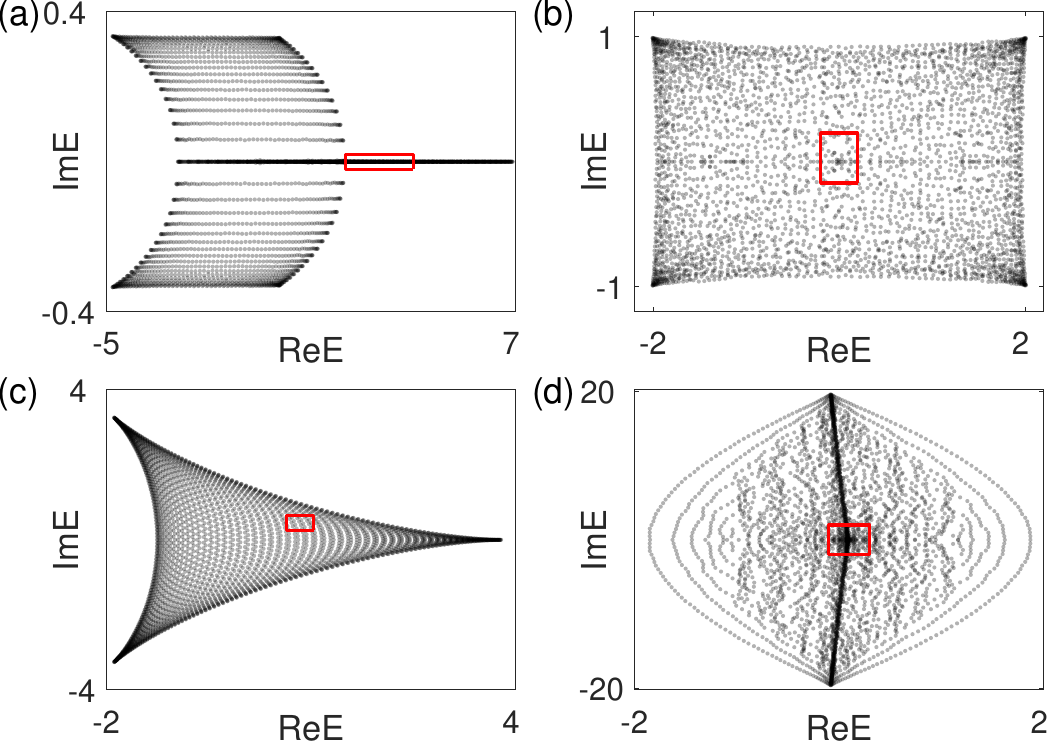}
\caption{Spectral stability and instability in the presence of different skin modes. Perturbed energy spectra for various models and geometries: (a) Model (\ref{fc_model}) with square geometry; (b) Model (\ref{gdse_model}) with rhombus geometry; (c) Model (\ref{jh_model}) with square geometry; (d) Model (\ref{am_model}) with square geometry. The system sizes are $N=3600,3613,3600,3600$, respectively. The bulk disorder strength is $\delta=0.2$ for all cases.}\label{fig15}
\end{figure}

In higher-dimensional non-Hermitian systems, spectral instability can also arise depending on the type of skin modes. Drawing insights from the 1D case, we expect spectral instability to occur in the presence of scale-free skin modes. We display the perturbed energy spectra when bulk disorder (with strength $\delta = 0.2$) is introduced for the four different cases in Fig. \ref{fig15}. These correspond to (a) model (\ref{fc_model}), square; (b) model (\ref{gdse_model}), rhombus; (c) model (\ref{jh_model}), square; and (d) model (\ref{am_model}), square. In the clean case, they exhibit normal skin modes, scale-free modes, anomalous corner skin modes, and a coexistence of normal and scale-free modes, respectively. For the normal and anomalous cases in (a) and (c), the energy spectra are nearly unchanged, signifying spectral stability. In contrast, for the critical NHSE in (b), significant changes in the spectra are observed compared to the clean counterparts shown in Fig. \ref{fig10}(b). The disorder blurs the spectral structure, making it resemble the spectra under the square geometry shown in Fig. \ref{fig10}(a). For the coexistence case in (d), the spectral region corresponding to scale-free modes is disrupted by disorder, while the central-line region corresponding to normal skin modes remains intact.

The numerical results indicate that scale-free modes are highly susceptible to perturbations, with instability occurring in the spectral region hosting these modes. We now provide an intuitive understanding of spectral stability/instability. In 1D, the normal NHSE is a stable phenomenon that can overcome Anderson localization for weak disorder \cite{jiangchen}. The normal skin modes have broadenings of order $O(1)$, and thus rarely “feel” the disorder. Similarly, in higher dimensions, the normal NHSE is also a stable effect, with skin modes having broadenings of order $O(1)$. These normal skin modes, as well as anomalous corner skin modes, are rarely affected by disorder, allowing them to withstand Anderson localization. The system requires a significant amount of disorder strength to destroy the spectral structure. In contrast, scale-free skin modes, with broadenings of order $O(L)$, are strongly influenced by disorder. Consequently, Anderson localization occurs even with small disorder strength, disrupting the spectral structure.

\subsection{Zero-perturbation limit vs TDL}\label{secviic}
We have demonstrated spectral instability arising from critical skin modes. These skin modes also yield the non-convergence of non-Bloch spectra in the TDL. Thus, spectral non-convergence and instability are linked through scale-free localization. It is intriguing to ask whether the perturbed spectra converge in the TDL. Let us denote 
\begin{eqnarray}
	\bar{\sigma}_G:~~\text{the perturbed spectra under geometry $G$}.
\end{eqnarray}
Spectral instability implies the non-exchangeability of the TDL and zero-perturbation limits. Therefore, the order in which these limits are taken (i.e., $\lim\limits_{L\rightarrow\infty}\lim\limits_{\delta\rightarrow 0}$ vs $\lim\limits_{\delta\rightarrow 0}\lim\limits_{L\rightarrow\infty}$) matters. In the following, we scrutinize these two limits under two scenarios: absence and presence of scale-free modes. In the first scenario, where the system lacks scale-free modes, the non-Bloch spectra converge and stabilize. The physical properties in the TDL for regular geometries are governed by our non-Bloch band theory. We have:
\begin{eqnarray}
\lim\limits_{L\rightarrow\infty}\lim\limits_{\delta\rightarrow 0}\bar{\sigma}_G =\lim\limits_{\delta\rightarrow 0}\lim\limits_{L\rightarrow\infty}\bar{\sigma}_G=\sigma_G.
\end{eqnarray} 

\begin{figure}[!t]
\centering
\includegraphics[width=3.375 in]{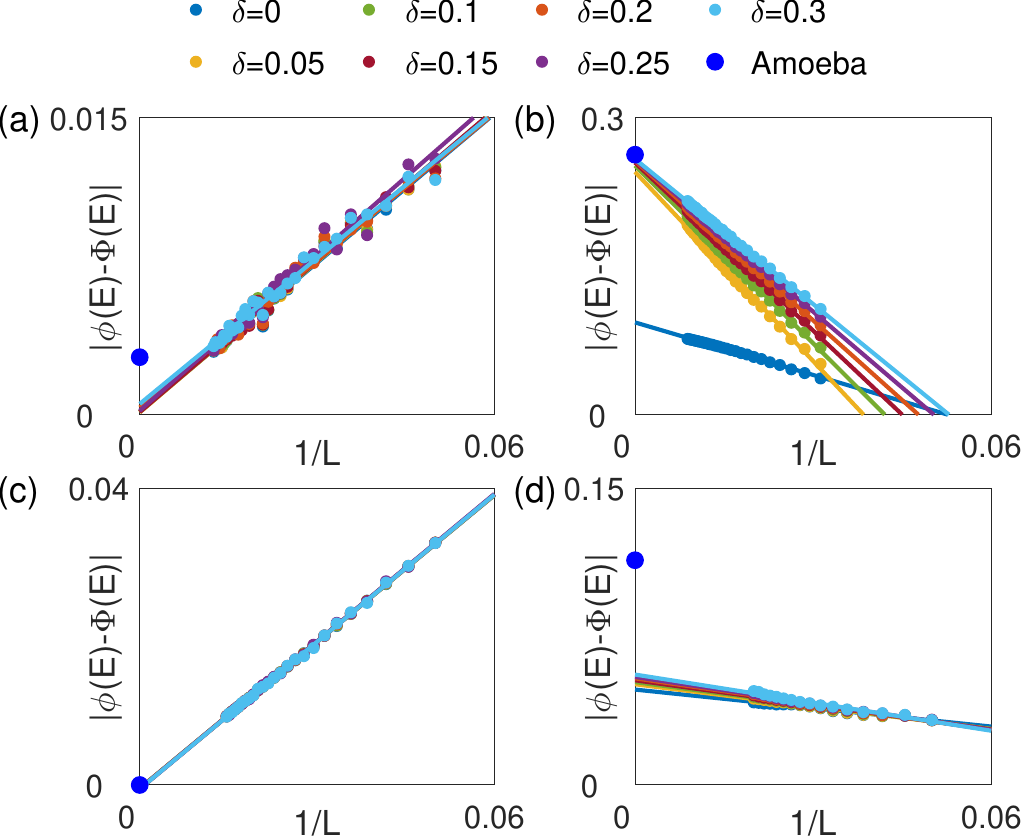}
\caption{Finite-size analysis of the spectral potential with varying disorder strength. The four panels correspond to the four marked regions in Fig. \ref{fig15}. The system possesses (a) normal skin modes, (b) scale-free skin modes, (c) anomalous corner skin modes, and (d) both the normal and scale-free skin modes. The vertical axis represents the absolute difference between the spectral potential obtained from numerical diagonalization ($\phi(E)$) and our formulation ($\Phi(E)$) in Eq. (\ref{2d_pot}). In each region, dozens of points are uniformly chosen and then averaged. The blue dots represent the values $|\Phi_{\text{Amoeba}}(E)-\Phi(E)|$, with $\Phi_{\text{Amoeba}}$ from the Amoeba formulation in Eq. (\ref{amoebap}).}\label{fig16}
\end{figure}
In the second scenario, where the system hosts scale-free modes, the first limit $\lim\limits_{L\rightarrow}\lim\limits_{\delta\rightarrow 0}$ is not well-defined as it pertains to the TDL for clean systems. For brevity, in the following, we consider the critical NHSE where all skin modes are scale-free. We have the following conjecture for the other limit.\\
\underline{\textbf{Conjecture-2:}}
\begin{eqnarray}
\lim\limits_{\delta\rightarrow 0}\lim\limits_{L\rightarrow\infty} \bar{\sigma}_{G} = \sigma_{\text{Amoeba}}. \label{relation6}
\end{eqnarray}
It asserts that in the TDL, the perturbed spectra become Amoeba spectra. While a rigorous proof seems impossible, we provide a physical argument. In the case of critical NHSE, the energy spectra for any regular lattice shapes become highly unstable and extremely sensitive to perturbations due to the presence of scale-free skin modes. Introducing disorder leads to Anderson localization, which stabilizes the spectra in the TDL. The disorder effectively erases the geometric information, making the spectra independent of any specific lattice shapes. In other words, the stabilization results in the geometry-irrelevant spectra $\sigma_{\text{Amoeba}}$.

We then present numerical evidence. For the four cases depicted in Fig. \ref{fig15}, we examine the spectral potential for four representative spectral regions (marked by red boxes in Fig. \ref{fig15}) with varying disorder strength $\delta$. The finite-size analysis of $|\phi(E) - \Phi(E)|$ is presented in Fig. \ref{fig16}. Here $\phi(E)$ and $\Phi(E)$ represent the spectral potential from exact diagonalization and our formulation. In each region, we uniformly select dozens of reference energy points and average their absolute deviations. For the normal and anomalous cases shown in Figs. \ref{fig16}(a)(c), it is evident that weak disorder barely alters the spectral potential, which converges to our theoretical value in the TDL. However, for the critical case shown in Fig. \ref{fig16}(b), the finite-size analysis indicates that $\lim\limits_{\delta \rightarrow 0}\lim\limits_{L \rightarrow \infty} \phi(E) = \Phi_{\text{Amoeba}}(E)$ in the presence of any weak disorder. Without disorder, taking the TDL yields another value, differing from both the Amoeba and our formulation. This coincides with the non-commutativity of $\delta\rightarrow 0$ and $L\rightarrow\infty$. For the coexistence case shown in Fig. \ref{fig16}(d), the normal/scale-free skin modes are stable/unstable against perturbation. The spectral potential deviates from both the Amoeba and our formulation. It is noteworthy that these conclusions apply universally across the selected spectral regions and number of points chosen inside.

\section{Classification}\label{secix}
Combining the previous discussions on non-Bloch band theory and spectral stability, we reach a complete classification of NHSE in arbitrary dimensions. This classification is comprehensive and exclusive. A detailed list of their properties is provided in Table \ref{table1}. There are two types of NHSE based on the net winding number criterion. Note that this classification does not account for any symmetries. The criterion enables a systematic discussion of the relationship between symmetry and the skin effect. For instance, with reciprocity symmetry \cite{yzsedge}, the net winding number along any direction is enforced to vanish, thus non-Hermitian systems with reciprocity symmetry are critical in the table. And the non-reciprocal NHSE is only compatible with certain point groups \cite{fangchen}.
\begin{table*}[!t]
\centering
\begin{tabular}{|c|c|c|c|c|c|}
\hline
Classification of NHSE & \multicolumn{4}{c|}{\textbf{Non-reciprocal}} & \textbf{Critical} \\
\hline
Criteria & \multicolumn{4}{c|}{$\exists j,~\bar{w}_j \neq 0$} & $\forall j,~\bar{w}_j = 0$ \\
\hline
Skin modes & normal & anomalous & boundary & \multicolumn{2}{c|}{scale-free}  \\
\hline
Spectral convergence in TDL? & \checkmark & \checkmark & \checkmark & \multicolumn{2}{c|}{$\times$} \\
\hline
Non-Bloch band theory & \checkmark & \checkmark & \checkmark & \multicolumn{2}{c|}{NA} \\
\hline
Does boundary ratio matter?  & $\times$ & $\times$ & $\times$ & \multicolumn{2}{c|}{\checkmark} \\
\hline
Stability against perturbation & \checkmark & \checkmark & \checkmark & \multicolumn{2}{c|}{$\times$} \\
\hline
Exchangeability of $\delta\rightarrow 0$ and TDL & \checkmark & \checkmark & \checkmark & \multicolumn{2}{c|}{$\times$} \\
\hline
$\lim\limits_{\delta\to0}\lim\limits_{L\to\infty}\bar{\sigma}_G$ & \multicolumn{4}{c|}{$\sigma_G$ in the absence of scale-free modes; unknown otherwise} & $\sigma_\text{Amoeba}$ \\
\hline
Example; Geometry & Eq. (\ref{cse_model}); \labeledsquare{} \labeledrhombus{} & Eq. (\ref{jh_model});\labeledsquare{} & Eq. (\ref{am_model});\labeledrhombus{}& Eq. (\ref{am_model});\labeledsquare{} & Eq. (\ref{gdse_model}); \labeledrhombus{} \\
\hline
\end{tabular}
\caption{Classification of NHSE in arbitrary dimensions. The classification criteria, types of skin modes (anomalous refers to anomalous corner skin modes), spectral convergence, applicability of the non-Bloch band theory, influence of boundary ratios on spectral structures, spectral stability under weak perturbation, exchangeability of zero-perturbation limit and the TDL, the perturbed spectra and representative models (geometries) are listed in each row. In the second row, $\bar{w}_j$ represents the net winding number along the $j$-th lattice-cut direction of the underlying geometric shape. In the fifth row, ``NA'' indicates not applicable due to the non-convergence of non-Bloch spectra. Within the table, ``$\times$" denotes ``No'' and ``$\checkmark$" denotes ``Yes''. In the nineth row, $\bar{\sigma}_G$, $\sigma_G$ and $\sigma_\text{Amoeba}$ denote the perturbed spectra, the non-Bloch spectra and the Amoeba spectra associated with the geometric shape $G$, respectively.}\label{table1}
\end{table*}

The first type is the critical NHSE (also dubbed reciprocal NHSE \cite{fangchen,yzsconjecture} in the literature), characterized by vanishing net winding numbers in all directions. The geometry-dependent skin effect described in the literature \cite{fangchen} falls into this category. For the critical NHSE, the eigenstates reside at the boundaries in a scale-free pattern. Due to this criticality, the energy spectra are highly sensitive to system size, boundary ratios, and weak perturbations. The spectral instability implies that the zero-perturbation limit and the TDL do not commute. Consequently, the non-Bloch spectra are not well-defined; however, any weak perturbations stabilize the spectra toward the Amoeba spectra $\sigma_{\text{Amoeba}}$.

The second type is the non-reciprocal NHSE. For a given geometry, there are certain lattice-cut directions along which the net winding number is nonzero. Note that the non-Bloch spectra may vary for different geometric shapes. The skin modes can manifest in various forms, including normal, anomalous, boundary, and scale-free skin modes. Different types of skin modes may coexist in a given geometry. Unless scale-free localization occurs, the non-Bloch bands are convergent in the TDL and stable against weak perturbations. They are fully captured by our non-Bloch band theory and potential formalism in Eqs. (\ref{2d_pot}) and (\ref{dd_pot}). The GBZ is well-defined for stable non-Bloch spectra. 

\section{Conclusions and discussions} \label{secx}
In conclusion, we developed a unified non-Bloch band theory for arbitrary-dimensional non-Hermitian systems. With the geometric information as input, we derived the spectral potential, DOS, and GBZ in the TDL for regular lattice geometries, highlighting their geometry-dependent nature. Regarding the NHSE, we systematically classified it into two types based on net winding numbers: critical and non-reciprocal. For the critical NHSE, we illustrated its scale-free skin modes, non-convergence of non-Bloch spectra, spectral instability in the presence of weak perturbations, and its relationship with the Amoeba formulation. For the non-reciprocal NHSE, we demonstrated various forms of skin modes, including normal, anomalous, boundary, and scale-free skin modes. We identified spectral convergence and stability in the absence of scale-free localization. Our framework establishes a solid foundation for comprehending non-Bloch bands, NHSE, their geometry-dependence and stability.

It is noteworthy that our formulation focuses on the dominant skin modes of order $O(L^d)$ in $d$D. Subleading modes, such as hybrid skin-topological modes \cite{llhhybrid,jbgong2022}, topological boundary states, and higher-order skin modes \cite{honhse1, honhse2,bergholtz2024b}, which are of order $O(L^j)(j<d)$, are not explicitly included in the spectral potential. How to effectively incorporate these subleading modes within the framework of non-Bloch band theory is an interesting question. In sharp contrast to Hermitian systems, where boundaries primarily affect boundary modes while leaving bulk modes intact, our findings highlight a significant distinction for higher-dimensional non-Hermitian systems: the geometric shape also influences the continuum non-Bloch bands in intriguing ways. Therefore,  our study hints the possibility of a novel bulk-edge-geometry correspondence. Specifically, topological boundary modes should be treated in a manner that accounts for geometric variations, and topological transitions may vary from shape to shape. We will leave the investigation of non-Bloch topological phases for future research.

Our work should inspire further exploration of high-dimensional non-Hermitian systems. We list several open questions. Firstly, while the two conjectures made in Eqs. (\ref{relation3}), (\ref{relation4}), and (\ref{relation6}) are supported by ample numerical evidence and physical arguments, a mathematically rigorous proof is highly desirable. Secondly, the emergence of scale-free localization fails a non-Bloch description solely based on lattice-cut directions due to spectral non-convergence in the TDL. It prompts the question: Can integrating additional information, such as boundary ratios, stabilize the non-Bloch spectra? If so, a non-Bloch band theory capable of handling scale-free skin modes may be developed, where different boundary ratios could yield distinct non-Bloch spectra in the TDL. According to the bulk-edge correspondence, topological boundary modes should also depend on these boundary ratios. Thirdly, how will various skin modes impact the transport and dynamic behaviors of higher-dimensional non-Hermitian systems? Fourthly, while a counterpart of 1D NHSE exists for open quantum systems \cite{lse1, lse2,wzlse,liuchunhui2020}, are there analogous Liouvillian skin effects in higher-dimensional open quantum systems? Finally, given the prevalence of non-Hermitian systems in photonic, acoustic, mechanical, and electrical platforms, as well as the feasibility of fabricating periodic lattice structures, we anticipate that our work will inspire further experimental studies. Specifically, we foresee investigations into the geometry-dependent spectral structures and the scale-free localization in the critical NHSE.

\begin{acknowledgments}
This work is supported by the National Key Research and Development Program of China (Grants No. 2023YFA1406704 and No. 2022YFA1405800) and the start-up grant of IOP-CAS.
\end{acknowledgments}

\appendix
\section{Proof of Eq. (\ref{1d_potential})}\label{appendixa}
In this appendix, we prove Eq. (\ref{1d_potential}) in Section \ref{seciib}. Leveraging the local form [See Eq. (\ref{1dlocal})] of the spectral potential in 1D, we need to prove:
\begin{eqnarray}
&\min_\mu& \int^{2\pi}_0 \frac{d k}{2\pi}\log|\det[H(e^{i k+\mu})-E]| \notag\\
&=&\sum_{j=p+1}^{p+q}\log|\beta_j(E)|+\log|f_q|,~\forall E.
\label{eqproof}
\end{eqnarray}
For a given reference enegry $E$, we sort the zeros of the ChP $f(\beta,E)=\det[H(\beta)-E]=\sum_{j=-p}^{q}f_j \beta^j$ as $|\beta_1|\leq|\beta_2|\leq\cdots\leq|\beta_p|\leq|\beta_{p+1}|\leq\cdots\leq|\beta_{p+q}|$. Note that $\log|(\cdot)|=\re\log[(\cdot)]$, the l.h.s. of Eq. (\ref{eqproof}) is
\begin{align}
	l.h.s.=\min_\mu \re\int^{2\pi}_0 \frac{d k}{2\pi}\; \log\det[(H(e^{ik+\mu})-E)].
\end{align} 
We then expand the $\log[(\cdot)]$ term 
\begin{align}\label{proofstep1}
&	\log\det[(H(e^{ik+\mu})-E)] \nonumber\\
&=\log[f_q (e^{ik+\mu})^{-p}\prod_{j=1}^{p+q}(e^{i k+\mu}-\beta_j)]\nonumber\\
	&=\log(f_q)-p\log(e^{i k+\mu}) \nonumber\\
	&+s\log(e^{i k+\mu})+\sum_{j=1}^{s}\log(1-\beta_j/e^{i k+\mu}) \nonumber\\
	&+\sum_{j=s+1}^{p+q}[\log(-\beta_j)+\log(1-e^{i k+\mu}/\beta_j)],
\end{align}
where we have assumed $|\beta_s|\leq e^{\mu}\leq|\beta_{s+1}|$. Upon Taylor-expanding all the logarithmics in Eq. (\ref{proofstep1}), the integrals of the $k$-dependent terms in the Taylor series vanish. We are left with
\begin{align}
	&	\log\det[(H(e^{ik+\mu})-E)] \nonumber\\
&=\log|f_q|+(s-p)\mu+\sum_{j=s+1}^{p+q}\log|\beta_j|,
\end{align}
where we have used $\re\log(\cdot)=\log|\cdot|$ again. There are three cases as per the value of $\mu$: (i) $s=p$; (ii) $s>p$ and (iii) $s<p$. We consider each case in the following.\\

\noindent (i) $s=p$. In this case, $s-p=0$, and we have
\begin{align}
	\int^{2\pi}_0 \frac{d k}{2\pi}\; \log|\det[H(e^{i k+\mu})-E]| \nonumber\\
	=\sum_{j=p+1}^{p+q}\log|\beta_j(E)|+\log|f_q|.
\end{align}
(ii) $s>p$.  In this case, we have $e^\mu\geq |\beta_j|$ for $p\leq j\leq s$, then
\begin{align}
	&\int^{2\pi}_0 \frac{d k}{2\pi}\; \log|H(e^{i k+\mu})-E| \nonumber\\
	&=\log|f_q|+\sum_{j=p+1}^{p+q}\log|\beta_j|+\sum_{j=p}^s (\mu-\log|\beta_j|) \nonumber\\
	&\geq \log|f_q|+\sum_{j=p+1}^{p+q}\log|\beta_j|.
\end{align}
(iii) $s<p$. In this case, we have $e^\mu\leq |\beta_j|$ for $s+1\leq j\leq p$, then
\begin{align}
	&\int^{2\pi}_0 \frac{d k}{2\pi}\; \log|H(e^{i k+\mu})-E| \nonumber\\
	&=\log|f_q|+\sum_{j=p+1}^{p+q}\log|\beta_j|+\sum_{j=s+1}^p (\log|\beta_j|-\mu) \nonumber\\
	&\geq \log|f_q|+\sum_{j=p+1}^{p+q}\log|\beta_j|;
\end{align}
Combining (i)(ii)(iii), we thus proved Eq. (\ref{eqproof}) and the equality holds only if $|\beta_p|\leq e^\mu\leq|\beta_{p+1}|$.

\section{Non-Bloch spectra of model (\ref{fc_model}) with square geometry}\label{appendixb}
\begin{figure}[!t]
\centering
\includegraphics[width=3.375in]{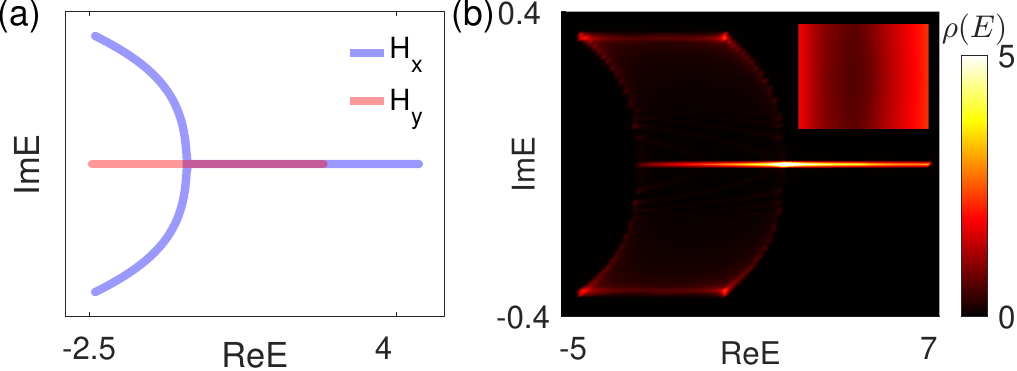}
\caption{Precise analysis of the non-Bloch spectra for model (\ref{fc_model}) under square geometry. (a) The 1D non-Bloch bands for the two components $H_x$ (blue) and $H_y$ (red) in the complex plane. (b) Exact spectral DOS extracted from Eq. (\ref{exactdos}). (Inset) Potential landscape $\Phi(E)$ as defined in Eq. (\ref{exactp}), whose Laplacian yields the DOS.}\label{figs1}
\end{figure}

In this appendix, we demonstrate how to analytically obtain the spectral potential and density of states (DOS) for model (\ref{fc_model}) on the square geometry. Since the $x$ and $y$ directions are decoupled, the Hamiltonian in the $(k_x,k_y)$ basis is written as $H(k_x,k_y)=H_x(k_x)+H_y(k_y)$. Let us perform the analytical continuation:
\begin{eqnarray}
H(\beta_x,\beta_y)=H_x(\beta_x)+H_y(\beta_y),
\end{eqnarray}
with $H_x(\beta_x)=\beta_x+\frac{3}{2}\beta_x^{-1}+\frac{1}{2}\beta_x^2+2\beta_x^{-2}$ and $H_y(\beta_y)=\frac{3}{2}\beta_y+\beta_y^{-1}$. The non-Bloch spectra for $H_x$ and $H_y$ can be obtained, respectively, as shown in Fig. \ref{figs1}(a). Since $H_y$ contains only nearest neighbor hoppings, its non-Bloch spectra reside on the real line. The linear spectral density is given by the following distribution \cite{hupotential}:
\begin{eqnarray}
\rho^{(y)}(E)=\frac{1}{\pi}\frac{1}{\sqrt{6-E^2}}.
\end{eqnarray}
The non-Bloch spectra on the square geometry is the superposition of those of $H_x$ and $H_y$. Intuitively, we may sweep the non-Bloch spectra of $H_x$ along the real axis, with the start and end points of this sweeping process being the two end points of the Bloch spectra of $H_y$. For $H_x$, its local spectral potential is given by Eq. (\ref{1dlocal}), i.e.,
\begin{eqnarray}
\Phi^{(x)}(E)=\log\frac{1}{2}+\log|\beta_3(E)|+\log|\beta_4(E)|.
\end{eqnarray}
Here $\beta_3$ and $\beta_4$ (which can be analytically obtained via Ferrari's root formula) are the third and fourth roots (sorted according to their moduli) of the ChP $H_x(\beta_x)-E=0$. The spectral potential generated by the 2D non-Bloch spectra is then
\begin{eqnarray}\label{exactp}
\Phi(E)=\int dE_y \rho^{(y)}(E_y) \Phi^{(x)}(E-E_y).
\end{eqnarray}
The spectral DOS is thus the Laplacian of $\Phi(E)$:
\begin{eqnarray}\label{exactdos}
\rho(E)=\frac{1}{2\pi}\int dE_y \rho^{(y)}(E_y) \nabla^2 \Phi^{(x)}(E-E_y).
\end{eqnarray}
This produces the exact spectral DOS in the TDL as shown in Fig. \ref{figs1}(b). It coincides with with the energy spectra from exact diagonalization and our formulation in Figs. \ref{fig7}(a,b) of the main text. We remark that the method above can be applied to any Hamiltonian with decoupled lattice-cut directions.

\section{Uniform spectra $\sigma_{\text{uniform}}$}\label{appendixc}
Similar to the Amoeba spectra $\sigma_{\text{Amoeba}}$, the uniform spectra $\sigma_{\text{uniform}}$ are irrelevant to geometric shapes. Compared to $\sigma_{\text{Amoeba}}$ defined from an algebraic analogy of the 1D GBZ condition, $\sigma_{\text{uniform}}$ is obtained from the perspective of point gaps. Point gaps are the topological origin of NHSE \cite{nhse6,huuniform}. In 1D, the PBC spectra form closed loops, whereas the OBC spectra form arcs embedded within these spectral loops. Introducing an imaginary gauge transformation (or inserting an imaginary flux) into the system, i.e., $H(k)\rightarrow H(k-i\mu)$, morphs the spectral loops, yet the OBC spectra remain intact. The OBC spectra stay inside the deformed spectral loops for any gauge transformation (or flux strength) \cite{nhse6}. Moreover, the OBC spectra are exactly the intersections:
\begin{eqnarray}\label{point} \sigma_{OBC}=\mathop{\bigcap}\limits_{\mu\in(-\infty,+\infty)} Sp(\mu). 
\end{eqnarray} 
Here, $Sp(\mu)$ denotes the region enclosed by the deformed spectral loop generated by flux strength $\mu$.

In $d>1$D, the uniform spectra are constructed by inserting an imaginary flux along each direction: 
\begin{eqnarray}\label{us} 
\sigma_{\text{uniform}}=\mathop{\bigcap}\limits_{\mu_j\in(-\infty,+\infty), j=1,2,...,d} Sp(\mu_1,\mu_2,\cdots,\mu_d).\notag\\
\end{eqnarray}
Here $Sp(\mu_1,\mu_2,\cdots,\mu_d)$ is the generalization of the rescaled spectra $Sp(\mu)$ in 1D. It contains the deformed Bloch spectra associated with flux strength $(\mu_1,\mu_2,\cdots,\mu_d)$ and their enclosed region. Formally, 
\begin{eqnarray}
Sp(\mu_1,\mu_2,\cdots,\mu_d):=\{E~|~\sum_{j=1}^d|w_j(E)|\neq 0\}.\label{pg}
\end{eqnarray} 
Here, $w_j(E)=\frac{1}{2\pi i}\int_{0}^{2\pi} \partial_{k_j}\log f(\beta_1,\beta_2,...,\beta_d,E)$ is the winding number along the $j$-th direction. The condition $\sum_{j=1}^d|w_j|\neq 0$ captures the intuitive notion that the NHSE can arise from the point gap in any direction. The OBC energy spectra are collapsed from these rescaled spectra until no spectral winding along any direction exists. This is realized by taking the intersections in Eq. (\ref{us}), which eliminates all possible spectral windings.

It is evident that $\sigma_{\text{uniform}}$ remains unchanged under basis transformations. In fact, a basis transformation would induce a shift in the rescaling factors and Eq. (\ref{us}) contains all possible rescaling factors. This implies the geometric independence of  $\sigma_{\text{uniform}}$. Furthermore, it can be proven that $\sigma_{\text{uniform}}=\sigma_{\text{Amoeba}}$ \cite{huuniform}. Since the OBC is unique in 1D, the non-Bloch spectra are exactly given by $\sigma_{\text{uniform}}$ or $\sigma_{\text{Amoeba}}$. However, in higher dimensions, the underlying geometric shape plays a crucial role in determining the structure (e.g., spectral range and DOS) of the non-Bloch spectra. This highlights a key difference between 1D and higher dimensions. When considering non-Bloch bands and NHSE in higher dimensions, the geometric information must be taken into account from the beginning.

%

\end{document}